\documentclass[entropy,article,submit,pdftex,moreauthors]{Definitions/mdpi} 
\firstpage{1} 
\pubvolume{1}
\issuenum{1}
\articlenumber{0}
\pubyear{2022}
\copyrightyear{2022}
\datereceived{} 
\dateaccepted{} 
\datepublished{} 
\hreflink{https://doi.org/} % If needed use \linebreak
\Title{Learned practical guidelines for evaluating Conditional Entropy and Mutual Information in discovering major factors of response-vs-covariate dynamics}

\TitleCitation{Learned practical guidelines for evaluating Conditional Entropy and Mutual Information in discovering major factors of response-vs-covariate dynamics}

\Author{Ting-Li Chen $^{1}$, Hsieh Fushing $^{2}$, and Elizabeth P. Chou $^{3,}$*}

\AuthorNames{Ting-Li Chen, Hsieh Fushing, and Elizabeth Chou}

\AuthorCitation{Chen, T.-L; Fushing, H.; Chou, E.}
\address{%
$^{1}$ \quad Institute of Statistical Science, Academia Sinica, Taipei 11529, Taiwan\\
$^{2}$ \quad Department of Statistics, University of California, Davis, CA 95616, USA\\
$^{3}$ \quad  Department of Statistics, National Chengchi University, Taipei 11605, Taiwan}

\corres{Correspondence:eptchou@g.nccu.edu.tw}

\abstract{We reformulate and reframe a series of increasingly complex parametric statistical topics into a framework of response-vs-covariate (Re-Co) dynamics that is described without any explicit functional structures. Then we resolve these topics' data analysis tasks by discovering major factors underlying such Re-Co dynamics by only making use of data's categorical nature. The major factor selection protocol at the heart of Categorical Exploratory Data Analysis (CEDA) paradigm is illustrated and carried out by employing Shannon's conditional entropy (CE) and mutual information ($I[Re; Co] $) as two key Information Theoretical measurements. Through the process of evaluating these two entropy-based measurements and resolving statistical tasks, we acquire several computational guidelines for carrying out the major factor selection protocol in a do-and-learn fashion. Specifically, practical guidelines are established for evaluating CE and $I[Re; Co] $ in accord with the criterion called [C1:confirmable]. Via [C1:confirmable] criterion, we make no attempts on acquiring consistent estimations of these theoretical information measurements. All evaluations are carried out on a contingency table platform, upon which the practical guidelines also provide ways of lessening effects of curse of dimensionality. We explicitly carry out six examples of Re-Co dynamics, within each of which, several widely extended scenarios are also explored and discussed.}

\keyword{Categorical exploratory data analysis; curse of dimensionality; Hierarchical clustering; interacting effects; K-means; LASSO.} 

\begin{document}

%%%%%%%%%%%%%%%%%%%%%%%%%%%%%%%%%%%%%%%%%%
%\setcounter{section}{-1} %% Remove this when starting to work on the template.
\section{Introduction}%1
Majority of scientific fields, such as biology\cite{faes}, neuroscience \cite{wibral}, medicine, sociology and psychology \cite{child} and many others \cite{contreas}, involve with dynamics of complex systems \cite{gellmann,adami}. Scientists and experts in such fields typically can only imagine or even brief outline various potential response-vs-covariate (Re-Co) relationships in an attempt to characterize dynamics of their complex systems of interest \cite{anderson}. Given no explicit functional form of such Re-Co relationships being available, such scientists still go ahead to collect structured data sets by investing great efforts in choosing which features for the role of response variable, and which features for the role covariate variables. Such choices of features are indeed critical for the sciences because their successes rely entirely on whether such structured data sets can embrace the essence of the targeted Re-Co dynamics or not.

Upon many successful scientific quests in aforementioned research areas, the targeted Re-Co dynamic rarely render an explicit system of equations, nor a complete set of functional descriptions. However, the data sets created by these successful scientists indeed are supposed to coherently reflect their curators' subject-matter knowledge and intelligence. From this perspective, the majority of data analysis on such structured data sets are tasked to decode curator's authentic knowledge and intelligence about the complex systems of interest under the setting of lacking of explicit functional forms of the targeted Re-Co dynamics.

In sharp contrast, nearly all statistical model-based data analyses on any structured data sets pertaining to wide-range of Re-Co dynamics always assume an explicit functional structure linking the response variables to covariate variables. Starting from hypothesis testing \cite{lehmann}, analysis of variance (ANOVA) to many variants of regression analysis\cite{fisher,scheffe}, including generalized linear models and log-linear models \cite{mccullagh,christensen}. By framing rather complex Re-Co dynamics with rather simplistic explicit functional structures, statistical model-based data analysis surely will run the dangers of hijacking data's authentic information content. With such dangers in mind, it is natural to ask the reverse question: {\bf What if we can reformulate all fundamental statistical tasks to fit under a framework of response-vs-covariate (Re-Co) dynamics without explicit functional forms, can we extract data's authentic information content of data sets?}

As the theme of this paper, we demonstrate a positive answer to the above fundamental question. The chief merits of such demonstrations are that we not only can basically do nearly all data analysis without statistical modeling, but more importantly we can reveal data's authentic information content to foster true understanding about the complex systems of interest. Our computational developments are illustrated through a series of 6 well-known statistical topic issues with increasing complexity.  All successfully revealed information content is visible and interpretable.

The positive answer resides in the paradigm called Categorical Exploratory Data Analysis (CEDA) with its heart anchored at a major factor selection protocol, which has been under developing in a series of published works \cite{FC21,FCC21, CCF22a, CCF22b} and a recently completed work \cite{FCC23}.  For demonstrating the positive answer, this paper establish practical guidelines for evaluating Theoretical Information Measurements, in particular Shannon's conditional entropy (CE) and mutual information between the response variables and covariate variables, denoted as $I[Re; Co]$ \cite{cover}, which are the basis of CEDA and major factor selection protocol.

Along the process of establishing such computational guidelines, we characterize four theme-components in CEDA and the major factor selection protocol:
\begin{description}
\item[[TC-1].] Our practical guidelines are established here for evaluating CE and $I[Re; Co] $ without requiring consistent estimations of their theoretical population-version of measurements.
\item[[TC-2].] All entropy-related evaluations are carried out on a contingency table platform, so learned practical guidelines also provide ways of relieving from effects of curse of dimensionality and ascertaining for [C1:confirmable] criterion, which is a kind of relative-reliability.
\item[[TC-3].] CEDA is free of man-made assumption and structures, so consequently its inferences are carried out with natural reliability.
\item[[TC-4].] CEDA only employs data's categorical nature, so the confirmed collection of major factors indeed reveals data's authentic information content disregarding data types.
\end{description}
The theme-component [TC-1] allows us to avoid many technical and difficult issues encountered in estimating the theoretical information measurement \cite{paninski,kraskov}. [TC-1] and [TC-2] together make CEDA's major factor selection protocol very distinct to model-based feature selection based on mutual information evaluations \cite{brown,vergara,bennasar,zhao}, while [TC-3] makes CEDA's inferences realistic, and [TC-4] makes CEDA to provide authentic information content with very wide applicability.

For specifically illuminating these four theme-components, we consider a structured data set consisting of data points that are measured and collected in a $L+K$D vector format with respect to $L+K$ features. The first $L$ components are the designated response (Re) features' measurements or categories, denoted as ${\cal Y}=(Y_1, ..., Y_L)'$, and the rest of $K$ components are $K$ covariate (Co) features' measurements or categories, denoted as $\{V_1,...,V_K\}$. It is essential to note that some or even all covariate features could be categorical. Thus, data analysts' task is prescribed as precisely extracting the authentic associative relations between ${\cal Y}$ and $\{V_1,...,V_K\}$ based on a structured data set.

By extracting authentic associations between response and covariate features, various Theoretical Information Measurements are employed under the structured data setting in \cite{FC21,FCC21, CCF22a, CCF22b, FCC23}. In particular, Re-Co directional associations developed in CEDA and its major factor selection protocol rely on evaluations of Shannon conditional entropy (CE) and mutual information ($I[Re; Co]$) that are all carried out upon the contingency table platform. This platform is indeed very flexible and adaptable to numbers of involving features on row- and column-axes as well as the total size of data points. Such a key characteristic makes CEDA very versatile in applicability. We explain in more detail as follows.

On the response side, a collection of categories of response features (pertaining to ${\cal Y}$) is determined with respect to their categorical nature and sample size. Likewise, on the covariate side, a collection of categories for each 1D covariate feature (pertaining to $V_k$ for $k=1, ..K$) is chosen accordingly. It is noted that a continuous feature is categorized with respect to its histogram \cite{hsiehroy}. If $L >1$, then the entire collection of response categories consists of all non-empty cells or hypercubes of $L$D contingency tables. Clustering algorithms, such as Hierarchical clustering or K-means algorithms, can be also performed for fusing $L$-dimensional response features into one single response variable. In regarding any fusing operations, the most basic key requirement is to retain the structural dependency among these $L$ response features. To this goal, both clustering algorithms are rather effective.

In contrast, singleton and joint (or interacting) effects of all possible subsets of $\{V_1,...,V_K\}$ are theoretically potential on the covariate side. However, it is practically known that any high order interacting effects needed to be considered are to a great extent determined by the sample size. That is, a covariate-vs-response contingency table platform can vary greatly in dimensions:  large or small. When viewing a contingency table as a high-dimensional histogram, which is a naive form of density estimation, the curse of dimensionality, or so-called finite sample phenomenon, is supposed to affect our conditional entropy evaluations whenever this table's dimension is large relative to data's sample size. We use the notation $C[A-vs-{\cal Y}]$  (rows-vs-columns) for a contingency table of a covariate variable subset $A\subseteq \{V_1,...,V_K\}$ and response variable ${\cal Y}$. As a convention, the categories of ${\cal Y}$ are arranged along its column-axis, while the categories of $A$ are arranged along the row-axis. This row-axis would expand with respect to memberships of $A$.

In CEDA, the associative patterns between any $A\subseteq \{V_1,...,V_K\}$ and ${\cal Y}$ would be discovered and evaluated upon the contingency table $C[A-vs-{\cal Y}]$. It is necessary to reiterate that $C[A-vs-{\cal Y}]$ can be viewed as a ``joint histogram'' or ``density estimation'' of all features contained in $A$ and ${\cal Y}$. From this perspective, when the dimension of $C[A-vs-{\cal Y}]$ increasingly expands as $A$ including more variables, it is expected that consequently its dimensionality would affect the comparability and reliability of conditional entropy evaluations.  Consequently, for comparability purpose, this criterion [C1:confirmable] in CEDA arises. This criterion is based on a so-called data mimicking operation developed in  \cite{FCC21}, as would be described as follows.

Let $\tilde{A}$ denote one mimicry of $A$ in the ideal sense of having the same deterministic and stochastic structures. In other words, $\tilde{A}$ is generated to have the same empirical categorical distribution of $A$, see \cite{FCC21} for construction details. More practically speaking, if the empirical categorical distribution of $A$ be represented by a contingency table, then, given the observed vector of row-sums, $\tilde{A}$ would be another contingency table that has the same lattice dimension and all its row-vectors are generated from Multinomial distribution with parameters specified by the corresponding row-sum and the corresponding vector of observed proportions in $A$'s contingency table. It is noted that $\tilde{A}$ is constructed independent of ${\cal Y}$, that is, $\tilde{A}$ is stochastically independent of ${\cal Y}$  \cite{FCC21}.

Denote the mutual information of ${\cal Y}$ of $A$ be $I[{\cal Y};A]$ based on $C[A-vs-{\cal Y}]$, and likewise $I[{\cal Y};\tilde{A}]$ based on $C[\tilde{A}-vs-{\cal Y}]$. The [C1:confirmable] used in CEDA is referred to the degree of certainty that $I[{\cal Y};A]$ is far beyond the upper limit of confidence region based on the empirical distribution of $I[{\cal Y};\tilde{A}]$. This [C1:confirmable] criterion indeed is in accord with CEDA's theme components: [TC-2] and [TC-3], regarding the merits of contingency table platform in dealing with curse of dimensionality and facilitating reliability. It is critical to note that we are not estimating the theoretical of mutual information of ${\cal Y}$ of $A$ here, and we just want to computationally make sure that $I[{\cal Y};A]$ is significantly above zero with great reliability under the reality of having only finite amount of data points in hand.

Henceforth, it is a critical fact in all applications of CEDA: a covariate feature set is confirmed as having effects on ${\cal Y}$ only when the [C1: confirmable] criterion of $I[{\cal Y};A]$ is established. This concept makes possible for [TC-1] by doing without the nonparametric estimations of Shannon entropy for a continuous distribution function as well as mutual information for two sets of continuous variables, which have been the long standing problems in physics and neural computing, see theoretical details in \cite{paninski} and computational protocols based on biGamma function in \cite{kraskov}.

Here, we do not take the view of contingency table as a setup of Grenander's Method of Sieves (MoS) \cite{grenander} in this paper. Though MoS can be a choice for practical reasons and computing issues involving many dimensional features or variables, we do not concern primarily on estimating the population-versions of CEs and $I[Re; Co]$ per se, nor the induced sieves biases. Rather, the dimensions of contingency tables are made adaptable to the necessity of accommodating multiple covariate feature-members in $A$. Within such cases, the collection of categories of $A$ might be built based on hierarchical or K-means clustering algorithms. From this perspective, computing for theoretical conditional entropy and mutual information between multiple dimensional covariate and possibly multi-dimensional $Y$ is neither realistically, not practically possible, due to limited sizes of available data sets. Since these kinds of sieves are data dependent. The computations for sieve biases can be much more complicate than that covered in \cite{paninski}.

In this paper, we illustrate and carry out CEDA coupled with its major factor selection protocol through a series of 6 classic statistical topic examples, within each of  which various scenarios are also considered. By building contingency tables across various dimensions with respect to different sample sizes, we attempt to reveal the robustness of CEDA resolutions to statistical topic issues. On one hand, we learn practical guidelines of evaluating conditional (Shannon) entropy and mutual information along this illustrative process. On the other hand, we demonstrate that very distinct CEDA resolutions to these classic statistical topic issues can be achieved by coherently extracting data's authentic information content, which is the intrinsic goals of any proper data analysis. That being said, if modeling is indeed a necessary step within a scientific quest, then data's authentic information content surely will serve its purpose better by relying on confirmed structures to begin with a new kind of data-driven modeling.

\section{Estimations of mutual information between one categorical and one quantitative variables.}
In this section, we demonstrate how to resolve classic statistical tasks by discovering major factors based on entropy evaluations.
First, we frame each classic statistical task into a precisely stated Re-Co dynamics. Secondly, we compute and discover major factors underlying this Re-Co dynamics. Inferences are then performed under [C1:confirmable] criterion across a spectrum of contingency tables with varying designed dimensions. Thirdly, we look beyond the setting of discussed examples to much wider related statistical topics.

Throughout this paper, all $95\%$ confidence ranges (CR) are calculated as the region between $2.5\%$ percentile on the lower tail and $97.5\%$ percentile on the upper tail of any simulated distribution. This CR reflecting both tail behaviors is considered informative. Since even when the upper tail is the only quantity of interest as being the case in this paper, the classic one-sided $97.5\%$ confidence interval becomes visible.

\subsection{[Example-1]: From 1D two-sample problem to one-way and two-way ANOVA. } Consider a data set consisting of quantitative observations  $\{Y_{lj}|l=1, 2; j=1,..., N_i\}$ of 1D response feature $Y$ derived from two populations labeled by $l=1, 2$, respectively. Let $Y_{lj}$ be distributed according to $F_l(.)$. Testing the distributional equality hypothesis ${\cal H}_0: F_1(y)=F_2(y),\; \forall y \in R^1$ is the most fundamental topic in statistics. Under this setting, the only covariate $V_1$ is the categorical population-ID taking values in $\{1,2\}$. The testing hypothesis problem and its subsequent ones can be turned into an equivalent problem: {\bf Is $V_1$ a major factor underlying the Re-Co dynamics of $Y$?} If $V_1$ is not a major factor, then ${\cal H}_0$ is accepted.  If ${\cal H}_0$ is indeed rejected by confirming $V_1$ being a major factor, then we would further want to discover where they are different.

For the illustrative simplicity, let $Y_{1j}\sim N(0, 1)$ and $Y_{1j}\sim N(1, 1)$ with $j=1,.., N/2$, that is,  $N_1=N_2$.  From the theoretical information measurement perspective, the theoretical value of entropy of $Y$ is calculated being equal to $H[Y]=1.5321$, and its conditional entropy
\[
H[Y|V_1]=(H[Y|V_1=0] +H[Y|V_1=1])/2=(1.4189 \times 2)/2=1.4189,
\]
so the mutual information shared by $Y$ and $V_1$ is denoted and calculated as $I[Y;V_1]=H[Y]-H[Y|V_1]=0.1132$. By $V_1$ being a major factor of $Y$, we mean that the $V_1$ is not replaceable by other covariate variables that is stochastically independent of $Y$, such as fair-coin-tossing random variable $\varepsilon $. That is, we theoretically establish this fact by knowing $0=I[Y;\varepsilon ] << I[Y;V_1]$.

In the real world, the two population-specific distributions $F_1(.)$ and $F_2(.)$ are often unknown. To accommodate this realistic setting, we build a histogram, say $\hat{F}(.)$, based on pooled observed dataset $\{Y_{ij}|i=1, 2; j=1,..., N_i\}$. With a chosen version of $\hat{F}(.)$ with $K'$ bins, we can build a $2\times K'$ contingency table, denoted by $C[V_1-vs-Y]$. Its two rows correspond to two population-IDs and all $K'$ bins with column-sums $n_k, k=1, ..K'$ being arranged along the column-axis. That is, $C[V_1-vs-Y]$ keeps the records of popultion-IDs for all members within each bin of $\hat{F}(.)$, and enable us to estimate the mutual information:
 \[
 I[Y; V_1]=H[Y]-H[Y|V_1]=H[V_1]-H[V_1|Y].
\]
All estimates of $I[Y; V_1]$ would be compared with estimates of $I[Y; \varepsilon]$ from $2\times K$ contingency tables generated as follows: its $k$th column with $k=1, ..,K'$ simulated from a binomial random variable $BN(n_k, P_{0})$ with $P_0=(N_1/N, N_2/N)'$. This comparison of
$I[Y; V_1]$ with $I[Y; \varepsilon]$ is a way of testing whether a major factor candidate satisfies the criterion [C1: confirmable] in \cite{CCF22a}. Precisely this testing is performed by comparing the observed estimate of $I[Y; V_1]$ with respect to the simulated distribution of $I[Y; \varepsilon]$.

To make our focal issue concrete and meaningful, we undertake the following simulation study, in which the reliability issue of $H[Y|V_1]$ estimation is addressed, and at the same time [C1: confirmable]  is tested. Recall that $Y_{1j}\sim N(0, 1)$ and $Y_{1j}\sim N(1, 1)$ with $j=1,.., N/2$. We consider two cases of $N=2000$ and $N=20,000$. For practical considerations with respect to the infinity range of Normality, we choose $K'=K+2$ bins for building a histogram via a $1+K+1$ fashion. The observed $90\%$ quantile range $[F_N^{-1}(0.05), F_N^{-1}(0.95) ]$ $K$ is divided into $K$ equal size of bins, while the first bin is $(-\infty, F_N^{-1}(0.05)]$ and last bin is $[F_N^{-1}(0.95), \infty)$. We use 5 choices of
$K\in \{10, 20, 30, 100, 1000\}$. For each $K$ value, the estimated Shannon entropy $H^{(K)}[Y]$ and conditional entropies $H^{(K)}[Y|V_1]$. Also, a $95\%$ confidence range (CR) of $I[Y; \varepsilon ]$ is also simulated and reported based on an ensemble of
$I^{(K)}[Y; \varepsilon]=H^{(K)}[Y]-H^{(K)}[Y|\varepsilon]$, where $\varepsilon$ is Bernoulli (fair-coin tossing) random variable.

 \begin{table}[]
 \centering
%\resizebox{\textwidth}{!}{%
\begin{tabular}{|l|lllll|}\hline
                N       & bin size & $H[Y]$ & $H[Y|V_1]$ & $I[Y; V_1]$ & $95\%$ CR of $I[Y; \varepsilon]$  \\\hline
\multirow{5}{*}{2000}  & 1+10+1    &  2.3993 &2.2824 &0.1168  &[0.00254, 0.00298]  \\
                       & 1+20+1   &3.0149& 2.8951 &0.1199    &[0.00489, 0.00551]  \\
                       & 1+30+1    & 3.3782 &3.2571 &0.1211 &[0.00757, 0.00836]       \\
                       & 1+100+1  &4.4424 &4.3043 &0.1382 &[0.02548, 0.02704]   \\
                       & 1+1000+1 &  6.2609 &5.9149 &0.3461 &[0.26435, 0.26768]     \\\hline
\multirow{5}{*}{20000} & 1+10+1  &2.4135 &2.3011 &0.1124   &[0.00025, 0.00030]   \\
                       & 1+20+1  & 3.0350 &2.9215 &0.1135 &[0.00050, 0.00057]     \\
                       & 1+30+1   & 3.3995 &3.2856 &0.1139  &[0.00074, 0.00082]    \\
                       & 1+100+1  &4.4807 &4.3649 &0.1157   &[0.00243, 0.00258]    \\
                       & 1+1000+1 &6.5310 &6.3933 &0.1377  &[0.02591, 0.02637]   \\\hline
\end{tabular}%
%}
\caption{Point estimations of mutual information $I[Y;V_1]$ with $0.1132$ as its theoretical value: $I[Y;V_1]=H[Y]-H[Y|V_1]=1.5321-1.4189$, and null $95\%$ confidence range (CR) of $I^{(K)}[Y; \varepsilon]$ with $\varepsilon$ being the Binomial random variable under the null hypothesis.}
\label{experiment1}
\end{table}

As reported in the table Table~\ref{experiment1}, it is evident that the mutual information $I^{(K)}[Y; V_1]=H^{(K)}[Y]-H^{(K)}[Y|V_1]$ is very close to the theoretical values as if they are nearly scale-free when $K=10, 20, 30$ with $N=2000$ and $K=10, 20, 30, 100$ with $N=20000$. The rule of thump in this 1D setting seems to be: the mutual information estimations are rather robust when the averaged cell count is over 30. When the average cell count is around 10, we begin to see the effects of finite sample phenomenon. Nonetheless, we still have estimates of $I^{(K)}[Y; V_1]$ being far above the upper limits of $95\%$ confidence range of $I[Y; \varepsilon ]$ when $K=100$ with $N=2000$ and even $K=1000$ with $N=20000$. This simulation indeed points to an observation that the conclusion based on $I^{(K)}[Y; V_1]$ tends to rather reliable in view of [C1: confirmable] criterion.

In summary, Table~\ref{experiment1} indicates that estimate of  mutual information of $I[Y|V_1]$ is far above the $95\%$ confidence range under the null hypothesis within each of all 5 choices of $K$ under the two cases of $N$. The 9 out of 10 cases have almost 0 p-values, except the $1+1000+1$ case with $N=2000$. These facts indicate one common observation: when all bins contain at least 20 data point, the estimate of $I[Y|V_1]$ is reasonably stably and practically valid. That is, we only need a stable and valid estimate of $I[Y|V_1]$ for the purpose of confirming a major factor candidacy.

In fact, it is surprising to see that, even when $K=1000$ in the case of $N=2000$, $I^{(K)}[Y; V_1]$ still retains [C1: confirmable] criterion by going beyond the upper limit of the $95\%$ confidence range of $I[Y; \varepsilon]$. This fact implies the correct decision still being retained because $V_1$ is confirmed as a major factor. These observations become crucial when estimations of $I[Y|V_1]$ are facing effects of curse of dimensionality, also called finite sample phenomenon.

As $V_1$ being determined as a major factor underlying the dynamics of $Y$ and the hypothesis ${\cal H}_0$ is rejected, we then can check which of $K+2$ bins' observed entropies fall inside or outside of bin-specific entropy-confidence-ranges built by simulated counts via $BN(n_k, P_{0})$ across $k=1, .., K+2$. By doing so, we discover where $F_1(.)$ and $F_2(.)$ are different locally.

Next, one very interesting observation is found and reported in Table~\ref{experiment1}: values of $H^{(K)}[Y]$ vary with respect to $K$, but $I^{(K)}[Y; V_1]$ is nearly scale-free (w.r.t $K$). We explain how this observation occurs. Let $f(y)=F'(y)$ be the hypothetical density function of random variable $Y$ with observed values $\{Y_{lj}|l=1, 2; j=1,..., N/2\}$. Based on fundamental theorem of calculus, for each $K$, we have the theoretical Shannon entropy $\tilde{H}(Y)$ is approximated as:
\begin{eqnarray*}
H[Y] &=& (-1) \int^{\infty}_{-\infty}f(y)\log{f(y)}dy,\\
&\cong& (-1) \sum^{K+1}_{k=0} f(y^*_k) \vartriangle(K) \log{f(y^*_k)},\\
&=& (-1) \sum^{K+1}_{k=0} p_k \log{\frac{p_k}{\vartriangle(K)}},\\
&=& H^{(K)}[Y] + \log{\vartriangle(K)},
\end{eqnarray*}
where $y^*_k$s denote inter-middle values in Mean Value Theorem of Calculus and $\vartriangle(K)=\frac{F_N^{-1}(0.95)-F_N^{-1}(0.05)}{K}$.

And we have
\[
\vartriangle(10)=J\vartriangle(J\times 10).
\]
with $J=2, 3, 10$ and $100$. Therefore, we have the approximating relations as:
\[
H^{(10)}[Y]\approx H^{(J\times 10)}[Y] - \log{J}.
\]
After some subtractions, the differences are close to $\log{2}$, $\log{3}$,
 $\log{10}$ and $\log{100}$, which matches with numbers shown in the 3rd column of Table~\ref{experiment1}.

By the same reason, these relations hold for estimated conditional entropies as well. That is, we also have: for all $K$s,
\[
H[Y|X] \cong H^{(K)}[Y|X] + \log{\vartriangle(K)},
\]
when all involving bins have 30 or so data points, as seen in 4th column of Table~\ref{experiment1}. This is the reason why that we see estimated values of $I^{(K)}[Y; V_1]$ being nearly constant (w.r.t $K$) when $K=10, 20, 30$ with $N=1000$ and $K=10, 20, 30, 100$ with $N=10,000$. This is a critical fact that we can employ mutual information estimates with reliability. Thus, we use the notation $I[Y; V_1]$ from here on, instead of $I^{(K)}[Y; V_1]$.

Here we further remark that the two-sample hypothesis testing problem ($L=2$) setting can be extended into the so-called multiple-sample problem ($L>2$) . Correspondingly, categorical variable $V_1$ of population-IDs is equipped with $L$ categories. This hypothesis testing:
 \[
 {\cal H}_0: F_l(y)=F(y),\; \forall y \in R^1, l=1, .., L.
 \]
retains the same equivalent formulation of as: {\bf Is $V_1$ a major factor underlying the dynamics of $Y$?} This multiple-sample problem is also known as one-way ANOVA, which is one fundamental topic problem in Analysis of Variance.

Another fundamental topic problem in Analysis of Variance is termed: two-way ANOVA, involving two categorical covariate features: $V_1$ and $V_2$. Let these two covariate features have $L_1$ and $L_2$ categories, respectively. Within a population with $V_1=l$ and $V_2=h$, measurements $Y_{lhj}$ are distributed with respect to $F_{lh}(.)$ with $l=1,.., L_1$ and $h=1, ..,L_2$.

The classic two-way ANOVA setting is specified by assuming Normality distribution $Y_{lhj}~N(\mu_{lh},\sigma^2)$ and $\mu_{lh}$ satisfying the following linear structure:
\[
\mu_{lh}=\mu+\alpha_l+\beta_h+\gamma_{lh},
\]
with $\mu$ as the overall effect, $\alpha_l$s the effects of $V_1$, $beta_h$s as effects of $V_2$, and $\gamma_{lh}$s as interacting effects of $V_1$ and $V_2$. These effects parameters are to satisfy the following linear constraints:
\[
\sum_{l=1} \alpha_l=\sum_{h=1} \beta_h=\sum_{l=1} \gamma_{lh}=\sum_{h=1} \gamma_{lh}=0.
\]
It is evident that this classic two-way ANOVA formulation is rather limited in the sense of excluding the possibility that $Y_{lhj}$ does not have an informative mean, such as non-normal distributions with heavy tails or more than one mode, or even lacking of the concept of mean, such as a categorical variable.

A much widely extended two-way version is given as follows:
\[
F_{lh}(.) \cong {\cal G}[{\cal M}_1 (V_1), {\cal M}_2 (V_2), {\cal M}_{12} (V_1, V_2)],
\]
where ${\cal G}[.]$ is unknown global function consisting of the following unknown component-wise mechanisms: the unknown component mechanism ${\cal M}_1 (V_1)$ having $V_1$ as its order-1 major factor; another unknown component mechanism ${\cal M}_2 (V_2)$ having $V_2$ as its order-1 major factor; and the unknown interacting component mechanism ${\cal M}_{12} (V_1, V_2)$ with $(V_1, V_2)$ as its order-2 major factor. Our goal of data analysis under this extended version is again reframed as computationally determining whether these order-1 and order-2 major factors are present or not underlying the Re-Co dynamics of $Y$ against the covariate features $V_1$ and $V_2$. If both covariate features $V_1$ and $V_2$ are independent or only slightly dependent with each other, the right major factor selection protocol can be found in \cite{CCF22a}. However, if they are heavily associated, an modified major factor selection protocol can be found in \cite{FCC23}.

We conclude this Example-1 with a summarizing statement: {\bf a large class of statistical topics can be rephrased and reframed into a major factor selection problem, and then this problem is resolved commonly by evaluating mutual information estimations that are not required to be precisely close to its unknown theoretical value.}

\subsection{[Example-2]: From dealing to lessening the effects of curse of dimensionality.}
It is noted here that, mutual information $I[Y; V_1]$ has another representation
\[
I[Y; V_1]=H[Y]+H[V_1]-H[Y, V_1]=\int_{R^2} dP(Y, V_1)\log\{\frac{ dP(Y, V_1)}{d(P(Y)\times P(V_1))}\}.
\]
This presentation is valid even for a categorical variable $V_1$. Based on this representation, we can clearly see the scale-free property of mutual information with respect to various choices of histograms. Nonetheless, we refrain from using this definition for estimating $I[Y; V_1]$. Since this definition-based estimation involves the estimation of joint distribution of $(Y, V_1)$, which is a harder problem due to its dimensionality. This so-called curse of dimensionality would become self-evident later on in our developments when the response variable ${\cal Y}$ and its covariate features $(V_1,..., V_k)$ are both multiple dimensional. The task of estimating multiple dimensional density become neither practical, nor reliable, given an ensemble of finite sample data points.

In this subsection, we demonstrate how to effectively deal with effects of curse of dimensionality. We consider again a two-sample problem, but having multiple dimensional data points, not single dimensional ones as in Example-1.  Again we denote two populations with IDs: $V_1=0$ and 1. Data points from these two populations are denoted as ${\cal Y}^{0}=(Y^{0}_1, .., Y^{0}_m)$ and ${\cal Y}^{1}=(Y^{1}_1, .., Y^{1}_m)$ with $m >1$, respectively. Let ${\cal Y}=(Y_1, .., Y_m)$ denote the multiple dimensional response variable. To resolve the same task of testing whether these two populations are equal with $m$ components possibly highly associative features, what would be the best way of building up the contingency table for the purposes of estimating the $I[{\cal Y}; V_1]$ for testing the hypotheses?

We expect that the equal-bin-size and equal-bin-area approaches for component-wise histograms are neither ideal nor practical due to curse of dimensionality. On the other hand, we know that the clusters of $m$-dim data points can naturally retain the dependency structures. Hence, it is intuitive to employ results of clustering algorithms to differentiate patterns of structural dependency within ${\cal Y}^{0}$ and ${\cal Y}^{1}$. This intuition leads to the important merit of cluster-based contingency table as a way of lessening effects from the curse of dimensionality. We illustrate these ideas through two samples of simulated multivariate Normal-distributed data described as follows.

Let $m=4$ and two mean-zeros Normal distributions: ${\cal Y}^{0} \sim N(\tilde{0}, \Sigma^{0})$ ${\cal Y}^{1} \sim N(\tilde{0}, \Sigma^{1})$.
\[
\Sigma^{0}=
  \begin{bmatrix}
    1 & \rho^0 & \rho^0 & \rho^0  \\
    \rho^0 & 1& \rho^0 & \rho^0 \\
    \rho^0 & \rho^0 & 1 & \rho^0\\
    \rho^0 & \rho^0 & \rho^0 & 1
  \end{bmatrix},
  \Sigma^{1}=
  \begin{bmatrix}
    1 & \rho^1 & \rho^1 & \rho^1  \\
    \rho^1 & 1& \rho^1 & \rho^1 \\
    \rho^1 & \rho^1 & 1 & \rho^1\\
    \rho^1 & \rho^1 & \rho^1 & 1
  \end{bmatrix}
\]
The Shannon entropies of these two 4D Normal distributions via the following formula with $d=4$:
\[
1/2 \log(det(\Sigma))+d/2(1+\log(2\pi))
\]
are calculated as 5.0942 and 4.4355, respectively. So the $H[Y|V_1]=(5.0942 + 4.4355)/2=4.7648$. As for $H[Y]$ of the mixture of two 4D Normal distributions, its calculation is not straightforward and even troublesome. Through an extra experiment using 100 millions of data points, we end with a negative estimate of the mutual information. This failed attempt in fact further provides a vivid clue of the effect of curse of dimensionality. In other words, we need to resolve such an effect by staying away from the rigid 4D hypercubes.

\begin{figure}
 \centering
 \includegraphics[width=6in]{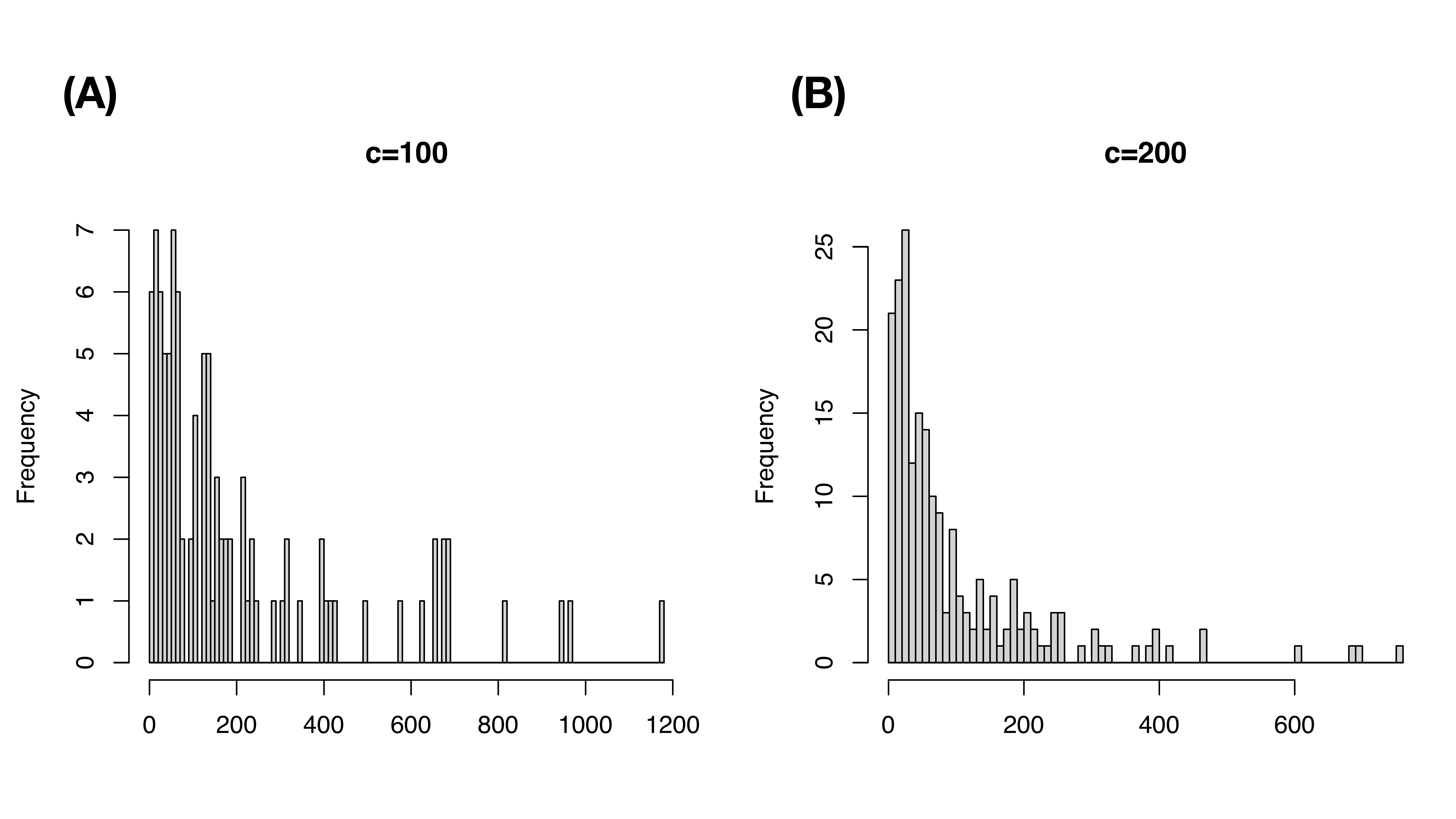}
 \caption{Comparing Hierarchical clustering and K-means via distributions of cluster sizes in Example 2: }
 \label{his2}
 \end{figure}

\begin{figure}
 \centering
 \includegraphics[width=6in]{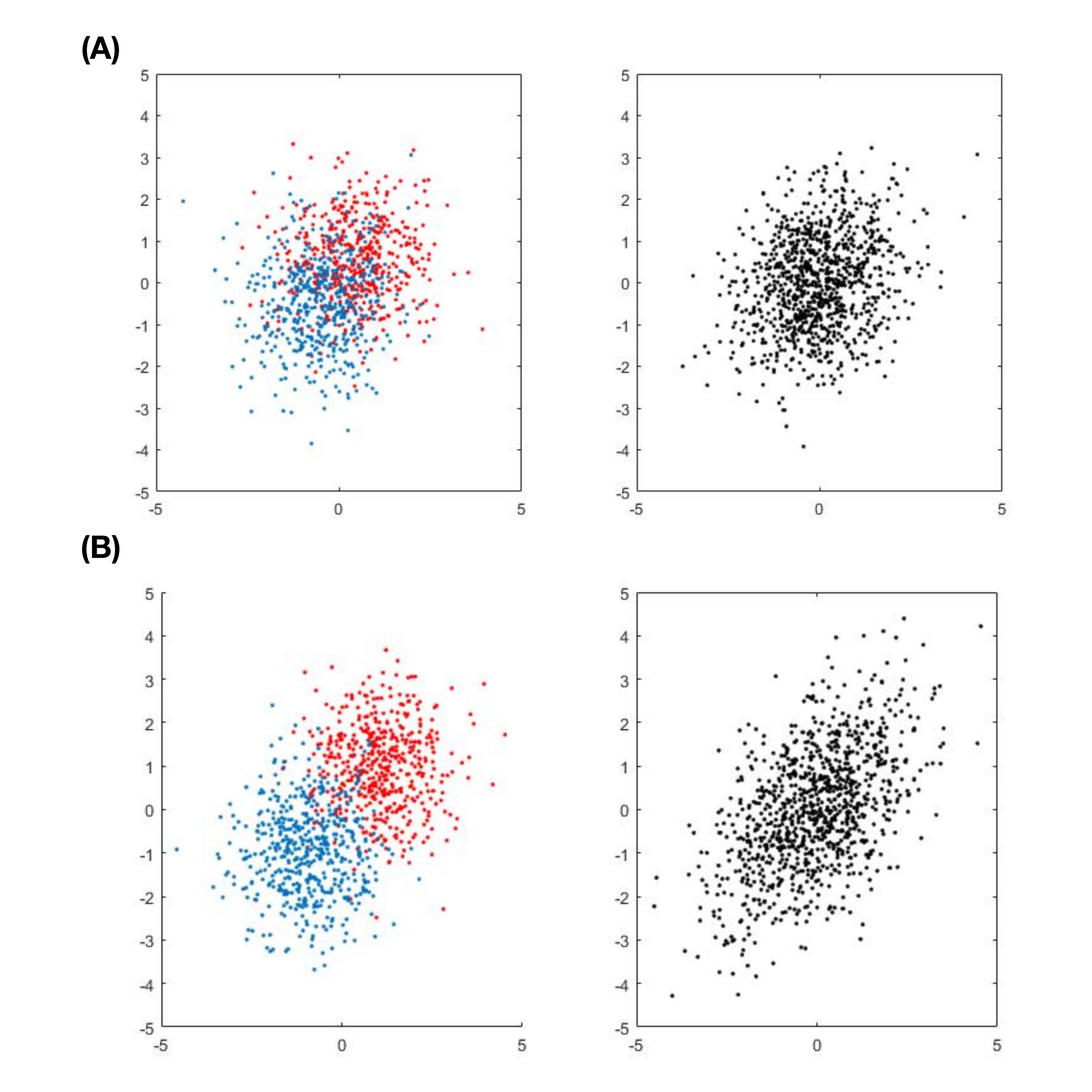}
 \caption{Two sets of pairwise scatter-plots of one simulated 2D normal mixture against 2D normal with equal mean vector and covariance matrix. The first set is for two close normal mixture with mean vectors: $(0.5 0.5)$ and $(-0.5,0.5)$ and the second is for relative apart normal mixture.}
 \label{fig:example2mix}
 \end{figure}

In contrast, we demonstrate that the cluster-based approaches are potentially reasonable choices to mend this effect of curse of dimensionality. Consider two commonly used clustering algorithms: Hierarchical clustering (HC) and K-means algorithms.  It is also known that the HC algorithm is computationally more costly than K-means algorithm. Since the HC-algorithm heavily relies on a distance matrix,  so HC-algorithm has difficulties in handling a data set with a very large sample size. Recently, very effective computing packages have been developed for K-means algorithm, that is, K-means algorithm can be effectively applied. On top of computing efficiency differences, there exists a critical difference between the two algorithms. The K-means provides much more even cluster-sizes than HC-algorithm does as illustrated in Figure~\ref{his2}, see also Figure~\ref{fig:example2mix}. For these reasons, we employ K-means clustering, not hierarchical clustering (HC), algorithm in the following series cases with $m=2, 3, 4$.

In this experiment, we take $\rho^0=0.5$ and $\rho^1=0.7$ under two settings with $N=2000$ and $N=20,000$. It is noted that the differences in $\rho^0$ values imply the differences in distribution shapes.  The series of clustering compositions are constructed as follows. We apply the K-means algorithm to derive a series of clustering compositions with 12, 22, 32 and 102 clusters. Correspondingly, we built a series of contingency tables of the formats: 1) $2 \times 12$; 2) $2 \times 22$; 3) $2 \times 32$ and 4) $2 \times 102$. With respect to the series of clustering compositions, we compute $H[Y]$ and $H[Y|V_1]$ and $I[Y; V_1]$. Here, $V_1$ is again the categorical variable of population-IDs.

\begin{figure}
 \centering
 \includegraphics[width=6in]{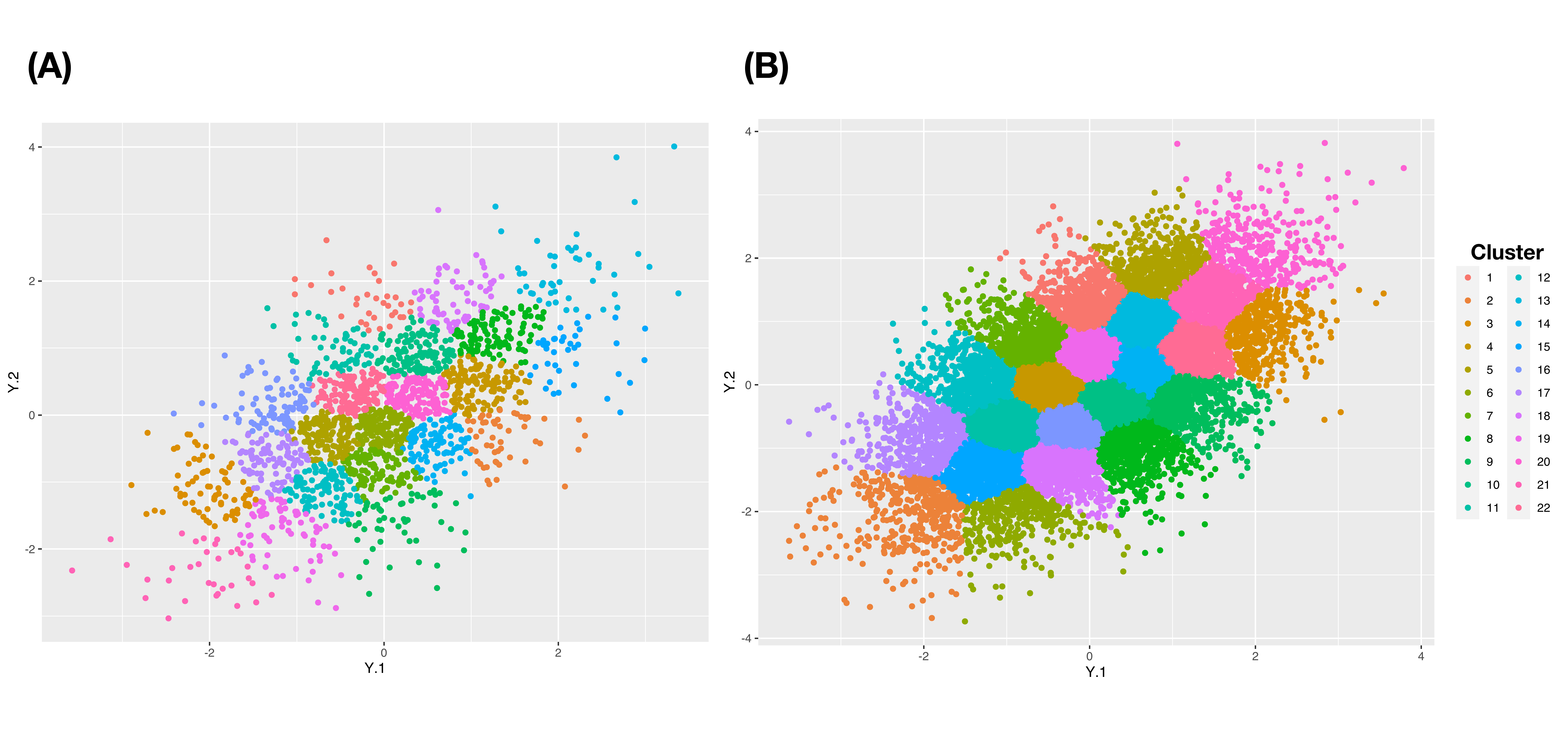}
 \caption{Kmean clusters in 2D setting: (A)$N=2000$; (B) $N=20,000$}
 \label{fig: example2mix}
 \end{figure}

\begin{table}[]
\centering
%\resizebox{\textwidth}{!}{%
\begin{tabular}{|l|lllll|}\hline
                n       & bin size & $H[Y]$ & $H[Y|X]$ & $I[Y; X]$ & $95\%$ CR of $I[Y; \varepsilon]$  \\\hline
\multirow{5}{*}{2000}  & 12    &2.3962 &2.3866& 0.0096&[0.00248, 0.00299] \\
                       & 22    &2.9722 &2.9530& 0.0192&[0.00487, 0.00544] \\
                       & 32     &3.3354 &3.3123 &0.0232&[0.00731, 0.00799]    \\
                       & 102   &4.5430 &4.4995 &0.0434 &[0.02576, 0.02711] \\
                       & 1002 &  6.7989 &6.4311 &0.3678&[0.33761, 0.34149]     \\\hline
\multirow{5}{*}{20000} & 12  &2.4208 &2.4148 &0.0060 &[0.00024, 0.00029]  \\
                       & 22  &  2.9916 &2.9816 &0.0100&[0.00049, 0.00056]  \\
                       & 32  & 3.3500 &3.3377& 0.0123&[0.00074, 0.00081]   \\
                       & 102 &4.5076 &4.4899& 0.0177&[0.00244, 0.00258]  \\
                       & 1002  &6.8662 &6.8236 &0.0425&[0.02570, 0.02608]  \\\hline
\end{tabular}%
%}
\caption{Entropies of Example-2 calculated from contingency tables built based on K-means clustering compositions on the 2D data setting.}
\label{experiment22D}
\end{table}

\begin{table}[]
\centering

%\resizebox{\textwidth}{!}{%
\begin{tabular}{|l|lllll|}\hline
                n       & bin size & $H[Y]$ & $H[Y|X]$ & $I[Y; X]$& $95\%$ CR of $I[Y; \varepsilon]$  \\\hline
\multirow{5}{*}{2000}  & 12    &2.4411 &2.4310 &0.0101&[0.00260, 0.00303]\\
                       & 22  &3.0166& 3.0028 &0.0138&[0.00476, 0.00537]\\
                       & 32   &3.3706 &3.3482 &0.0224 &[0.00732, 0.00812]  \\
                       & 102  &4.5297& 4.4771 &0.0526&[0.02563, 0.02712]\\
                       & 1002&6.8065 &6.4558 &0.3507 &[0.33899, 0.34254]     \\\hline
\multirow{5}{*}{20000} & 12  &2.4642& 2.4620 &0.0023 &[0.00025, 0.00030] \\
                       & 22 &3.0425 &3.0337& 0.0088&[0.00047, 0.00053] \\
                       & 32  &3.4064 &3.3958 &0.0106&[0.00075, 0.00083] \\
                       & 102    &4.5307 &4.5067& 0.0241&[0.00246, 0.00258]   \\
                       & 1002  &6.8551 &6.7988& 0.0563&[0.02582, 0.02632]  \\\hline
\end{tabular}%
%}
\caption{Entropies of Example-2 calculated from contingency tables built based on K-means clustering compositions on the 3D data setting.}
\label{experiment23D}
\end{table}

\begin{table}[]
\centering
%\resizebox{\textwidth}{!}{%
\begin{tabular}{|l|lllll|}\hline
                n       & bin size & $H[Y]$ & $H[Y|X]$ & $I[Y; X]$ & $95\%$ CR of $I[Y; \varepsilon]$  \\\hline
\multirow{5}{*}{1000}  & 12    &2.4599 &2.4556 &0.0043 & [0.00249, 0.00299] \\
                       & 22    &3.0612& 3.0518& 0.0094  &[0.00477, 0.00536] \\
                       & 32     &3.4115& 3.3911& 0.0204  &[0.00753, 0.00838]    \\
                       & 102   &4.5065& 4.4508& 0.0557  &[0.02565, 0.02717] \\
                       & 1002 &  6.8162& 6.4627& 0.3535  &[0.33696, 0.34110]     \\\hline
\multirow{5}{*}{10000} & 12  &2.4756& 2.4728& 0.0029  &[0.00026, 0.00032]  \\
                       & 22  &  3.0772& 3.0736& 0.0036  &[0.00049, 0.00056]  \\
                       & 32  & 3.4456& 3.4377& 0.0079  &[0.00073, 0.00081]   \\
                       & 102 &4.5590& 4.5347& 0.0243 &[0.00244, 0.00257]  \\
                       & 1002  &6.8328 &6.7697 &0.0631&[0.02556, 0.02607]  \\\hline
\end{tabular}%
%}
\caption{Entropies of Example-2 calculated from contingency tables built based on K-means clustering compositions on the 4D data setting.}
\label{experiment24D}
\end{table}

The messages derived from Example-1 are also observed in Example-2 across 2D to 4D settings in Table~\ref{experiment22D}, Table~\ref{experiment23D} and Table~\ref{experiment24D}. These results clearly indicate that distribution shape differences can be effectively and reliably picked up by entropy-based evaluations of mutual information between the $Y$ and categorical label variable $V_1$. These results imply that we widely extend one-way ANOVA and two-way ANOVA settings to accommodate high dimensional data points as we have argued in Example-1.

In order to better understand the limit of such entropy-based approach, we twist the 2D setting in Example-2 a little bit. This more complicate version of Example-2, denoted as Example-$2^*$, consists of one 2D normal mixture and one 2D normal. These two 2D distributions are further made to have equal mean vector and covariance matrix. Furthermore, two kinds of mixture-settings are designed and used. The first setting of Example-$2^*$ is designed for a mixture of two relatively close 2D normal with mean vectors: $(0.5 0.5)$ and $(-0.5,0.5)$. The second setting is designed for relatively apart normal mixture with mean vectors: $(-1,-1)$ and $(1,1)$.. These two settings of pairwise scatter-plots are given in Figure~\ref{fig:example2mix}. It is obvious that we can visually separate the two 2D distributions in the second mixture setting, but can not do equally well in the first mixture setting.

\begin{table}[]
\centering
%\resizebox{\textwidth}{!}{%
\begin{tabular}{|l|lllll|}\hline
                data       & bin size & $H[Y]$ & $H[Y|X]$ & $I[Y; X]$& $95\%$ CR of $I[Y; \varepsilon]$  \\\hline
\multirow{5}{*}{1st mixture}  & 12    &2.4246& 2.4233 &0.0012 &[0.00258, 0.00309] \\
                       & 22    & 2.9958 &2.9910& 0.0048&[0.00506, 0.00575] \\
                       & 32      &3.3805& 3.3725& 0.0080 &[0.00786, 0.00855]   \\
                       & 102    &4.5481& 4.5214& 0.0267&[0.02622, 0.02747] \\
                       & 1002  & 6.7953 &6.4700& 0.3252&[0.33811, 0.34153]   \\\hline
\multirow{5}{*}{2nd mixture} & 12   &2.4434& 2.4375& 0.0059 &[0.00226, 0.00272]  \\
                       & 22   &2.9943& 2.9795& 0.0147 &[0.00529, 0.00602]  \\
                       & 32   & 3.3678 &3.3518& 0.0159&[0.00745, 0.00817]   \\
                       & 102 & 4.5485 &4.5143& 0.0342&[0.02542, 0.02690] \\
                       & 1002   & 6.7975 &6.4573& 0.3403&[0.33702, 0.34059]  \\\hline
\end{tabular}%
%}
\caption{Entropies of two settings of Example-$2^*$ calculated from contingency tables built based on K-means clustering compositions with $N=20,000$.}
\label{experiment23}
\end{table}

The mutual information estimates and confidence ranges under the null hypothesis are calculated and reported in Table~\ref{experiment23}. In the first mixture setting, it is apparent that $V_1$ fails to be a major factor by failing to satisfy the criterion [C1: confirmable] across all $K$ choices. This result is coherent with our visualization through the upper panel Figure~\ref{fig:example2mix}. As for the 2nd mixture setting, $V_1$ is claimed as a major factor by satisfying the [C1: confirmable] criterion across all $K$ choices. This result is also coherent with our visualization through the lower panel Figure~\ref{fig:example2mix}. Further, we observe that the relative position of $I[Y; X]$ estimates against upper and lower limits of null confidence ranges are rather stable when the sizes of clusters are not too small. This observation indeed provides us the practical guideline for varying choices of $K$ according to different sample sizes when we employ mutual information to perform inferences under Re-Co dynamics.

We conclude this Example-2 (Example-$2^*$)  with a summarizing statement: {\bf Though, any theoretical evaluations of mutual information under the presence of high dimensionality are practically impossible, clustering algorithms provide practical guidelines for building contingency tables and evaluating mutual information for inferential purposes by lessening the effects of curse of dimensionality.}

\subsection{[Example-3]: From linear to highly nonlinear associations.}
We then turn to consider the simplest one-sample problem involving dependent 2D data points. The framework of Re-Co dynamics is self-evident. In this example, we examine the validity and performances of inferences based on estimated mutual information between two 1D continuous random variables $Y$ and $X$ via contingency tables of various dimensions. For simplicity in the first scenario of Example-3, we consider bivariate normal $(Y, X) \sim N(\tilde{0}, \Sigma)$ with covariance matrix:
\[
\Sigma=
  \begin{bmatrix}
    1 & \rho \\
    \rho & 1
  \end{bmatrix},
\]
Here the correlation coefficient $\rho$ is taken to be $0.0$ and $0.5$, respectively, in this experiment with $N=1000$ or $10,000$. The contingency tables are derived from K-means algorithm being applied on $X$ and $Y$, respectively, with a series of pre-determined numbers of clusters: $\{12, 22, 32, 102\}$.

For the setting of $\rho=0$, we report the calculated $I[Y; X]$ and confidence range of $I[Y; \varepsilon]$ in Table~\ref{experiment30} across the 16 dimensions of contingency tables. The smallest size of contingency table has $144 (=12\times 12)$ cells. Its averaged cell-count is less than 14 for $N=2000$. The largest size of contingency table is $102\times 102$, which is more than $10^4$. Its averaged cell-counts is less than 2 for $N=20000$.

From the upper half of Table~\ref{experiment30} for the $N=2000$, all estimates of  $I[Y; X]$ are beyond the upper limit of $95\%$ confidence range of $I[Y; \varepsilon]$. That is, the hypothesis of $Y$ and $X$ being independent is falsely rejected.  In contrast, from the lower half of Table~\ref{experiment30} for the $N=20000$, all estimates of  $I[Y; X]$ are either below the lower limit of $95\%$ confidence interval of
$I[Y; \varepsilon]$ or within confidence range, except the results based on the largest $102\times 102$ contingency table. That is, the same independency hypothesis would be not be falsely rejected except in the case of the largest contingency table. Such a contrasting comparison between the upper and lower halves of Table~\ref{experiment30} clearly indicates that validity of mutual information evaluations heavily rely on degrees of volatility of cells counts, especially on testing independence. We further explicitly express such volatility below.

A simple reasoning for the above results goes as follows. For this independent setting of $Y$ and $X$, for expositional simplicity, let all cells in contingency tables have equal probability. In the smallest contingency table, the cell probability is $1/144$. The cell-count is a random variable with mean and variance being very close to $N/144$ as well. Thus, the cell-count is falling between $N/144 \pm 2\sqrt{N/144}$ with at least $95\%$. With $N=2000$, the $95\%$ range is close to $[6, 22]$, while with $N=20000$ the $95\%$ range is close to $[110, 150]$. Based on these two $95\%$ intervals, we can see that the Shannon entropy along each row of $12 \times 12$ contingency table can be volatile with $N=2000$, while it is not the case with $N=20000$. In fact, when $N=2000$, a $6\times 6$ contingency table indeed provides much more stable evaluations of mutual information.

\begin{table}[]
\centering
%\resizebox{\textwidth}{!}{%
\begin{tabular}{|l|lllll|}\hline
                bin size       & bin size & $H[Y]$ & $H[Y|X]$ & $I[Y; X]$&  $95\%$ CR of $I[Y; \varepsilon]$ \\ \hline
\multirow{4}{*}{Y=12} & X=12   &2.4135 &2.3435 &0.0700& [0.0637, 0.0669]\\
& X=22&2.4135 &2.2861 &0.1273& [0.1231, 0.1274]\\
& X=32 &2.4135 &2.2194 &0.1940& [0.1863, 0.1916]\\
& X=102 &2.4135 &1.7971 &0.6164& [0.5714, 0.5787]\\\hline
\multirow{4}{*}{Y=22} & X=12   &3.0168& 2.9014 &0.1154& [0.1249, 0.1294]\\
& X=22&3.0168& 2.7650& 0.2517 &[0.2393, 0.2450]\\
& X=32&3.0168 &2.6319 &0.3848& [0.3613, 0.3681]\\
& X=102 &3.0168 &2.0360 &0.9808 &[0.9365, 0.9439]\\\hline
\multirow{4}{*}{Y=32} & X=12   &3.3910 &3.1952 &0.1958 &[0.1899, 0.1951]\\
& X=22&3.3910& 3.0196 &0.3714& [0.3587, 0.3656]\\
& X=32&3.3910& 2.8494 &0.5416& [0.5143, 0.5209]\\
& X=102 &3.3910& 2.1175 &1.2736& [1.2040, 1.2106]\\\hline
\multirow{4}{*}{Y=102} & X=12   &4.5236 &3.9131 &0.6105& [0.5657, 0.5728]\\
& X=22&4.5236 &3.5339 &0.9897& [0.9516, 0.9585]\\
& X=32&4.5236& 3.2717 &1.2519 &[1.2193, 1.2261]\\
& X=102 &4.5236 &2.2962 &2.2274 &[2.1571, 2.1643]\\\hline\hline

\multirow{4}{*}{Y=12} & X=12   &2.3392 &2.3332 &0.0059& [0.0060, 0.0063]\\
& X=22&2.3392 &2.3275& 0.0116& [0.0115, 0.0119]\\
& X=32&2.3392 &2.3216 &0.0175 &[0.0172, 0.0177]\\
& X=102 &2.3392 &2.2799 &0.0592& [0.0578, 0.0588]\\\hline
\multirow{4}{*}{Y=22} & X=12   &2.9424 &2.9311 &0.0113& [0.0116, 0.0120]\\
& X=22&2.9424 &2.9215 &0.0210& [0.0223, 0.0228]\\
& X=32&2.9424 &2.9109 &0.0316 &[0.0335, 0.0342]\\
& X=102 &2.9424 &2.8334 &0.1090& [0.1122, 0.1135]\\\hline
\multirow{4}{*}{Y=32} & X=12  &3.3311 &3.3155 &0.0157& [0.0174, 0.0179]\\
& X=22&3.3311 &3.2978 &0.0333 &[0.0334, 0.0341]\\
& X=32&3.3311 &3.2843& 0.0468 &[0.0496, 0.0505]\\
& X=102  & 3.3311 &3.1634 &0.1677& [0.1675, 0.1690]\\\hline
\multirow{4}{*}{Y=102} & X=12   &4.5504 &4.4933 &0.0571& [0.0582, 0.0592]\\
& X=22&4.5504 &4.4401& 0.1103& [0.1116, 0.1128]\\
& X=32&4.5504 &4.3836 &0.1668 &[0.1684, 0.1698]\\
& X=102  & 4.5504 &3.9991&0.5513& [0.5475, 0.5497]\\\hline
\end{tabular}%
%}
\caption{$(Y, X) \sim MN((0,1), \Sigma)$ with $\rho=0.0$ and $ N=2000 $ (upper half) , $n=20000$(lower half).}
\label{experiment30}
\end{table}

%***********
In the setting of $\rho=0.5$, we report the calculated $I[Y; X]$ and confidence range of $I[Y; \varepsilon]$ in Table~\ref{experiment31} across the 16 dimensions of contingency tables with $N=20000$. We observe that the calculated $I[Y; X]$ is far above the upper limit of confidence interval of $I[Y; \varepsilon]$ even in the largest contingency table with dimension $102\times 102$. The reason is that the number of effectively occupied cells are much smaller due to the dependency, that is, many cells supposed to be empty are indeed empty. With many empty cells coupling with many occupied cells with relatively large cell counts, the Shannon entropy is evaluated with great stability. These results from independent and dependent experimental cases are learned to constitute practical guidelines for evaluating mutual information.

\begin{table}[]
\centering
%\resizebox{\textwidth}{!}{%
\begin{tabular}{|l|lllll|}\hline
                bin size       & bin size & $H[Y]$ & $H[Y|X]$ & $I[Y; X]$&  95\% CR of $I[Y; Z]$ \\ \hline
\multirow{4}{*}{Y=12} & X=12  & 2.3317 &2.1839 &0.1478 &[0.0058, 0.0062]\\
& X=22  &2.3317 &2.1758 &0.1559 &[0.0114, 0.0119]\\
& X=32  &2.3317 &2.1709 &0.1609 &[0.0175, 0.0180]\\
& X=102  &2.3317 &2.1270 &0.2048 &[0.0578, 0.0588]\\\hline
\multirow{4}{*}{Y=22} & X=12  &2.9543 &2.7995 &0.1548 &[0.0116, 0.0120]\\
& X=22 &2.9543 &2.7852 &0.1692 &[0.0224, 0.0230]\\
& X=32 &2.9543 &2.7750 &0.1793 &[0.0336, 0.0344]\\
& X=102 &2.9543 &2.7018 &0.2525 &[0.1125, 0.1139]\\\hline
\multirow{4}{*}{Y=32} & X=12 &3.3654 &3.2043 &0.1611 &[0.0172, 0.0178]\\
& X=22 &3.3654 &3.1864 &0.1790 &[0.0332, 0.0339]\\
& X=32 &3.3654 &3.1708 &0.1945 &[0.0492, 0.0501]\\
& X=102 &3.3654 &3.0555 &0.3099 &[0.1672, 0.1688]\\ \hline
\multirow{4}{*}{Y=102} & X=12 &4.5415 &4.3416 &0.1999 &[0.0583, 0.0590]\\
& X=22 &4.5415 &4.2849 &0.2565 &[0.1117, 0.1131]\\
& X=32 &4.5415 &4.2344 &0.3070 &[0.1654, 0.1668]\\
& X=102 &4.5415 &3.8806 &0.6609 &[0.5488, 0.5513]\\\hline
\end{tabular}%
%}
\caption{$(Y, X) \sim MN(\tilde{0}, \Sigma)$ with $\rho=0.5$ and $N=20000$}
\label{experiment31}
\end{table}

\begin{figure}
 \centering
 \includegraphics[width=6in]{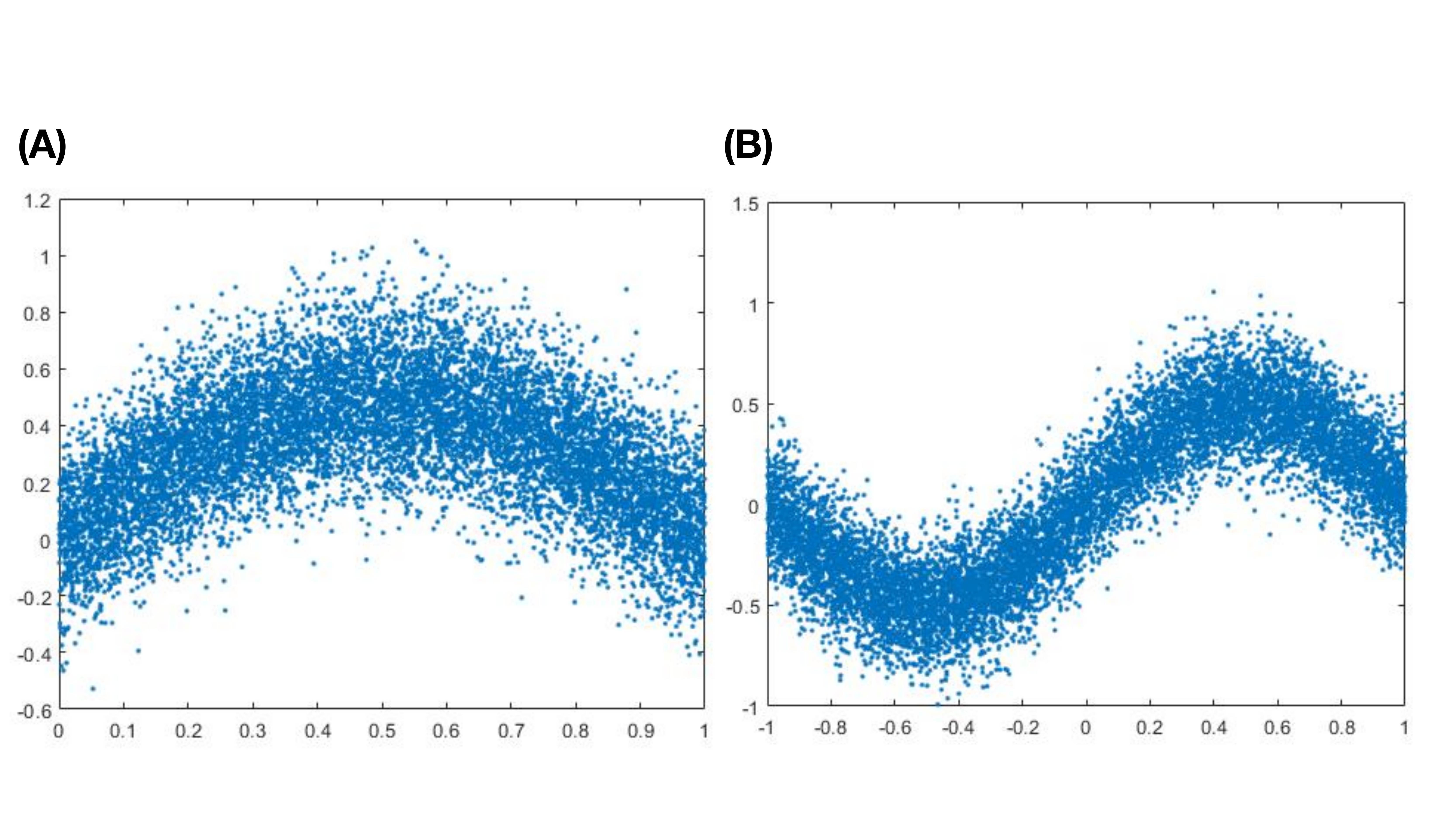}
 \caption{Two scatter-plots of two simulated data sets in sine functional shapes.}
 \label{fig:example3sine}
 \end{figure}

The second scenario of Example-3 is about whether the calculated mutual information $I[Y; X]$ can reveal the existence of non-linear association between $Y$ and $X$. We generate two simulated data sets based on two non-linear associations: 1) half-sine function; 2) full-sine function, as shown in the two panels of Figure~\ref{fig:example3sine}. Within both cases of non-linear associations, it is noted that the correlations of $Y$ and $X$ are basically equal to zero.

\begin{table}[]
\centering
%\resizebox{\textwidth}{!}{%
\begin{tabular}{|l|lllll|}\hline
                bin size       & bin size & $H[Y]$ & $H[Y|X]$ & $I[Y; X]$& $95\%$ CR of $I[Y; \varepsilon]$  \\\hline
\multirow{4}{*}{Y=12}  & X=12  &  2.4840 &1.7450 &0.7391 &[0.00600, 0.00632]\\
 & X=22  & 2.4840 &1.7326 &0.7514 &[0.01137, 0.01183]\\
 & X=32  & 2.4840& 1.7237 &0.7603 &[0.01690, 0.01742]\\
 & X=102  & 2.4840& 1.6977&0.7863 &[0.05704, 0.05804]\\\hline
\multirow{4}{*}{Y=22} & X=12   &3.0881 &2.3131 &0.7749 &[0.01151, 0.01189]\\
& X=22  &3.0881 &2.2975 &0.7906 &[0.02205, 0.02264]\\
& X=32  &3.0881 &2.2853 &0.8028& [0.03305, 0.03387]\\
& X=102  & 3.0881 &2.2369 &0.8512 &[0.11254, 0.11387]\\\hline
\multirow{4}{*}{Y=32} & X=12   &3.4499& 2.6679& 0.7819& [0.01689, 0.01743]\\
& X=22  &3.4499& 2.6466 &0.8033& [0.03272, 0.03343]\\
& X=32  &3.4499& 2.6335 &0.8163& [0.04928, 0.05009]\\
& X=102  & 3.4499& 2.5559  &0.8940& [0.17143, 0.17303]\\\hline
\multirow{4}{*}{Y=102} & X=12   &4.6133 &3.7972& 0.8161& [0.05663, 0.05757]\\
& X=22  &4.6133 &3.7550& 0.8583& [0.11237, 0.11375]\\
& X=32  &4.6133& 3.7194 &0.8939 &[0.17072, 0.17235]\\
& X=102  & 4.6133 &3.5164 &1.0969 &[0.56831, 0.57063]\\\hline

\end{tabular}%
%}
\caption{Evaluations of entropy, conditional entropy and mutual information under the half-sine simulation study.}
\label{experiment32}
\end{table}

In the setting of half-sine functional relation, we report the calculated $I[Y; X]$ and confidence range of $I[Y; \varepsilon]$ in Table~\ref{experiment32} across the 16 dimensions of contingency tables with $N=20000$. Across all 16 dimensions of contingency tables, the calculated $I[Y; X]$ are far beyond the upper limits of confidence intervals of $I[Y; \varepsilon]$. As far as p-value being concerned, they are all basically zeros. The same results are observed in the setting of full-sine functional relation as reported in Table~\ref{experiment32}. These two settings in this non-linear association scenario together demonstrate that the calculated $I[Y; X]$ can reveal the existence of significant association between $Y$ and $X$. This demonstration is important in the sense of without knowing the functional forms of their association.

\begin{table}[]
\centering
%\resizebox{\textwidth}{!}{%
\begin{tabular}{|l|lllll|}\hline
                bin size       & bin size & $H[Y]$ & $H[Y|X]$ & $I[Y; X]$& $95\%$ CR of $I[Y; \varepsilon]$  \\\hline
\multirow{4}{*}{Y=12}  & X=12  &  2.4807 &2.1916 &0.2890& [0.0061, 0.0064]\\
& X=22  &2.4807 &2.1822 &0.2984& [0.0115, 0.0120]\\
& X=32  &2.4807 &2.1757& 0.3050& [0.0171, 0.0176]\\
& X=102  & 2.4807 &2.1310 &0.3497& [0.0567, 0.0577]\\\hline
\multirow{4}{*}{Y=22} & X=12   &3.0692 &2.7651 &0.3042& [0.0114, 0.0118]\\
& X=22  &3.0692 &2.7517& 0.3175 &[0.0223, 0.0229]\\
& X=32  &3.0692 &2.7426 &0.3266 &[0.0333, 0.0341]\\
& X=102  & 3.0692 &2.6671 &0.4022& [0.1133, 0.1147]\\\hline
\multirow{4}{*}{Y=32} & X=12   &3.4398& 3.1293 &0.3105 &[0.0170, 0.0175]\\
& X=22  &3.4398 &3.1094 &0.3303 &[0.0331, 0.0338]\\
& X=32  &3.4398 &3.0980 &0.3417 &[0.0493, 0.0502]\\
& X=102  & 3.4398 &2.9917& 0.4481 &[0.1717, 0.1735]\\\hline
\multirow{4}{*}{Y=102} & X=12   &4.5698 &4.2185 &0.3513 &[0.0577, 0.0587]\\
& X=22  &4.5698 &4.1752& 0.3946 &[0.1118, 0.1130]\\
& X=32  &4.5698 &4.1233 &0.4466 &[0.1679, 0.1694]\\
& X=102  & 4.5698 &3.7851& 0.7848 &[0.5577, 0.5602]\\\hline

\end{tabular}%
%}
\caption{Evaluations of entropy, conditional entropy and mutual information under the whole-sine simulation study.}
\label{experiment33}
\end{table}

We summarize what practical guidelines we learn from Example-1 through Example-3 in this section. The most apparent fact is that the calculated values of mutual information $I[Y; X]$ vary with respect to dimensions of contingency tables. However, the good news is that the amounts of variations are relative small and even very minute when cell-counts in the contingency table are not too low. Nonetheless, that the calculated mutual information $I[Y; X]$ is very capable of revealing the presence and absence of associations underlying Re-Co dynamics of response variable $Y$ and covariate variable $X$ from the three examples and scenarios therein considered in this section. And it is a reliable way of seeking consistent inferential decisions by varying contingency tables' dimensions. This capability can be made very efficient if we choose the dimension of contingency table to suitably reflecting the total sample size of data set with varying degrees. That is, we make sure such efficiency is achieved by varying the dimensions of contingency tables from small to reasonably large. The final guideline is that comparability between two mutual information evaluations is resting on their more or less identical computational platforms, that is, their contingency tables are more or less the same in dimensions. On the other hand, the averaged numbers of cell counts are relatively large, mutual information evaluations are rather robust to some degrees of differences in contingency tables' dimensions. These practical guidelines will ascertain mutual information evaluations always coupled with reliability. Finally, the data-types of $Y$ and $X$ are entirely free because we rely on the their categorical nature only.

\section{Examples with complex Re-Co dynamics.}
Next, we consider two examples with more complex Re-Co dynamics than the three examples discussed in the previous section. Through these two examples of having independent covariate features, we further illustrate the necessity of following practical guidelines motivated and learned in the previous section.
\subsection{[Example-4]: From complex interaction to further beyond.}
After going through three relative simple examples in the previous section, we now turn to examples with more complex Re-Co dynamics. Consider a functional relation between $Y$ and $\{X_1, ..X_4\}$ specified as follows:
\[
Y=X_1+\sin(2\pi (X_2+X_3))+N(0,1)/10
\]
with $\{X_1, ..X_4\}$ being i.i.d. $U[0,1]$ and $N=10,000$. That is, $X_4$ plays the role of observable noise random variable, while unobservable noise is $N(0,1)/10$. Our goal is to discover the order-1 major factors $X_1$ and order-2 major factor $(X_2, X_3)$. It is worth noting that this order-2 major factor can not be discovered via linear regression analysis, even when the product type of interacting effect is included in the model.

The response variable $Y$ is categorized with 12 bins, so does each of the 4 covariate features. We calculate mutual information of $Y$ and all possible feature subsets' $A \subseteq \{X_1, ..X_4\}$, say $I[Y; A]$.  If $|A|=k$, we build a $(12)^k \times 12$ contingency table for calculating for evaluating $I[Y; A]$. Here $A$ also stands for a fused categorical variable in the sense that categories of $A$ are all occupied $k$D hypercubes of its $k (=|A|)$ feature-members.

We compute and report conditional entropies (CEs) for all possible $A$s and arrange them with respect to sizes $|A|$ of $A$ in Table~\ref{summary1}. Also we report a term called successive (S) CE-drops defined via the following CEs difference:
\[
SCE_{drop}[Y|A]=(H[Y]-H[Y|A]) -\max_{A'\subset A}\{H[Y]-H[Y|A']\}=\min_{A'\subset A}\{H[Y|A']\}-H[Y|A].
\]
This SCE term is designed to evaluate the extra effect of CE-drop by including an extra feature-member. The above formula is precise in theory. But in reflecting the aforementioned last practical guideline in the last section, it is essential to note that $SCE[Y|A]$ involves at least two different settings of $|A|=k$ and $|A'| =k' (< k)$, which correspondingly involve two different dimensions of contingency tables: one is of $(12)^k \times 12$ and the other is $(12)^{k'} \times 12$. Therefore, based on what we have learned from the previous section, these settings renders different scales of conditional entropy and mutual information computations. That is, these different scales will certainly make mutual information evaluations not completely comparable, especially when cell-counts in the contingency tables are overall too small. For instance,
\[
SCE_{drop}[Y|X_1, X_2]=0.0644=H[Y|X_1]-H[Y|X_1, X_2]=2.2315-2.1671.
\]

The SCE-drop of $(X_1,X_2)$ is more than 10 times of CE-drop of $X_2$. It would be a mistake to claim that $X_1$ and $X_2$ are conditional dependent given $Y$. Since the scale in evaluating $H[Y|X_1]$ is different from the scale in evaluating $H[Y|X_1X,_2]$. Nevertheless, since $X_4$ plays a role of random noise in this example, the information contents of $X_1$ and $(X_1, X_4)$ are supposed to be very close from the perspective of their contingency table. Theoretically, we have $H[Y|X_1]=H[Y|X_1, X_4]$. That is, $H[Y|X_1, X_4]$ should represent the information content of $X_1$ upon the setting of $(12)^2 \times 12$ contingency table. Along this line of argument, we should refine the SCE-drop as follows:
\[
SCE^*_{drop}[Y|X_1, X_2]=H[Y|X_1, X_4]-H[Y|X_1, X_2]=2.1685-2.1671=0.0014.
\]
Via the same argument, this SCE-drop should be compared with $H[Y|X_4]-H[Y|X_2, X_4]=2.4557-2.3780=0.0777$, which is 5 times larger than $0.0014$. Hence, it is obvious that $X_1$ and $X_2$ do not have joint interacting effects. In fact, it would be more precise evaluation of the effect of $X_2$ under the 2-feature setting if we use $H[Y|X_4, X_5]-H[Y|X_2, X_4]$ with $X_5$ being another irrelevant independent $U[0,1]$ random variable. However, according to the guidelines learned from example-1 and -2, $H[Y|X_4, X_5]$ and $H[Y|X_4]$ should be relatively close because of the sample size $10,000$.

This line argument ultimately converges to the following practical guideline on evaluating Information Theoretical measurements via contingency table platform: ``these CEs and mutual information measurements are comparable only when they are evaluated under the same dimensions of contingency tables''. This guideline indeed is coherent with a statistical concept of conditioning with respect to the observed row-sum vector.

Before summarizing our findings from Table~\ref{summary1}, where we reported calculated CEs and $SCE_{drop}$, we need to prepare baseline-evaluations to make sure that all CEs comparisons are sensible. Here, we recall that $C[A-vs-Y]$ denotes for the contingency table with categories of $Y$ on column-axis and categories of covariate feature subset $A$ on row-axis.
\begin{description}
\item[1-feature setting:] With $C[X_1-vs-Y]$ having its proportion vector of row-sums denoted as $P_{X_1}$, we build an ensemble of $C[X^{\varepsilon}_1-vs-Y]$ by distributing $i-$th column-sum $N[Y=i]$ with respect to $Multinomial(N[Y=i], P_{X_1})$. The average of CEs of $H[Y| X^{\varepsilon}_1]$, denoted as ${\cal E}[H[Y| X^{\varepsilon}_1]]$  is designed to be comparable with $H[Y|X_1]$. Their difference ${\cal E}[H[Y|X^{\varepsilon}_1]]-H[Y|X_1]$ is a proper and valid measurement of the CE-drop of $X_1$. Likewise for each of the rest covariate features.
\item[2-feature setting:] With $C[Y; (X_1, X_2)-vs-Y]$, we need to compute ${\cal E}[H[Y|(X_1, X_2)^{\varepsilon}]]$ for the joint CE-drop of
$(X_1, X_2)$ calculated as ${\cal E}[H[Y|(X_1, X_2)^{\varepsilon}]]-H[Y|(X_1, X_2)]$. We also need
${\cal E}[H[Y|(X_1, X^{\varepsilon}_2)]]$ for calculating $SCE^*_{drop}[Y|X_1, X_2]$ in order to be able to compare to
 ${\cal E}[H[Y|(X_1, X_2)^{\varepsilon}]]-{\cal E}[H[Y|(X^{\varepsilon}_1, X_2)]]$ to figure out the amount $I[(X_1, X_2)|Y]-I[(X_1, X_2)]$.

As for $(X_2, X_3)$, in comparisons with SCEs of $(X_2, X_4)$ and  $(X_3, X_4)$, its $SCE_{drop}$ is calculated as $0.7781$, which is more than 10 times of $X_3$'s individual $SCE_{drop}$. This is a very strong indication of interacting effect of $(X_2, X_3)$ due to evident presence of their conditional dependency given $Y$. This fact establishes the feature-pair $(X_2, X_3)$ as an order-2 major factor.

\item[3-feature setting:] In Table~\ref{summary1}, the $SCE_{drop}$ of feature-triplet $(X_1, X_2, X_3)$ from feature-pair $(X_2, X_3)$ is $0.8431$, which is about 3.5 times of CE-drop of $X_1$. This observation could seemingly point to the potential presence of conditional dependency of $(X_1, X_2, X_3)$. However, if we more precisely calculate the effect of $X_1$ when adding to $(X_2, X_3)$ as:
\[
SCE^*_{drop}[Y|X_1, X_2, X_3]=H[Y|X_2, X_3, X_4]-H[Y|X_1, X_2, X_3]=1.2263-0.8362=0.3901,
\]
and compare it with $H[Y|X_4, X_5, X_6]-H[Y|X_1, X_4, X_5]$ with $X_5$ and $X_6$ being independent random variables, which is expected to be larger than $0.2322$, but small than $0.3901$. Therefore, we can only confirm the ecological effect does exist between $X_1$ and $(X_2, X_3)$, that is, they can be order-1 and order-2 major factors of $Y$. But, certainly they don't form conditional dependency underlying $Y$, see details of major factor selection protocol in \cite{CCF22a}.
\end{description}

\begin{table}[]
\centering
\resizebox{\textwidth}{!}{%
\begin{tabular}{llllllllllll}\hline
1Feature & CE	& SCE-drop	 &2Feature   &CE	      & SCE-drop &3Feature	   & CE	       & SCE-drop	& 4Feature	&CE  & SCE-drop  \\ \hline
X1 &	2.2315 &	0.2322       & X1\_X2	& 2.1671 &	0.0644    &X1\_X2\_X3&	0.8362 &	0.8431	& X1\_X2\_X3\_X4 &	0.1762	& 0.6599\\
X2 &	2.4579 & 0.0057        & X1\_X3	& 2.1647 & 0.0667     &X1\_X2\_X4&	1.4451 &	0.7219	&	                         &             & \\
X3 &	2.4575 &	0.0062       & X1\_X4	& 2.1685 & 0.0630     &X1\_X3\_X4&	1.4531 & 0.7115	&		                     &             & \\
X4 &	2.4557 &0.0079         & X2\_X3	& 1.6793 &	0.7781    &X2\_X3\_X4&	1.2263 & 0.4530	&	                          &            &  \\
	 &                 &              	  & X2\_X4	& 2.3780	& 0.0777	  &				   &              &              &                           &            &  \\
	 &		             &                  & X3\_X4	& 2.3831 & 0.0726	  &				   &              &              &                           &            &  \\\hline

\end{tabular}%
}
\caption{Experiment with $Y=X_1+\sin(2\pi (X_2+X_3))+N(0,1)/10$ and $N=10,000$. Each categorized 1-features has 12 bins, so a $k$-feature has $(12)^k$ $k$D hypercubes.}
\label{summary1}
\end{table}

\subsection{[Example-5]: From high-order interaction to complexity.}
In order to see the effect of higher order major factor,  we change the functional form of $Y$ slightly as:
\[
Y=X_1+\sin(2\pi (X_2+X_3+X_4))+N(0,1)/10.
\]
With sample size $N=10,000$, our computational results are reported in Table~\ref{summary2}. Likewise, we can confirm $X_1$ as an order-1 major factor and triplet $(X_2, X_3, X_4)$ as an order-3 major factor. In sharp contrast, the evidence of order-3 major factor seems to disappear when $N=1000$, as shown in Table~\ref{summary3}. This is the exact demonstration of the effect of finite sample phenomenon, or curse of dimensionality. Do these two contrasting results: presence and absence of order-3 major factor in $N=10,000$ and $N=1000$, respectively, mean that we should give up in looking for high order major factors on small data sets?

\begin{table}[]
\resizebox{\textwidth}{!}{%
\begin{tabular}{llllllllllll}\hline
1Feature & CE	& CE-drop	 &2Feature   &CE	      & SCE-drop &3Feature	   & CE	       & SCE-drop	& 4Feature	&CE  & SCE-drop  \\ \hline
X1 &	2.2299 &	0.2295      & X1\_X2	& 2.1636 &	0.0662    &X1\_X2\_X3&	1.4444 &	0.7191	& X1\_X2\_X3\_X4 &	0.1945	& 1.0367\\
X2 &	2.4539 & 0.0055        & X1\_X3	& 	2.1671 & 0.0627     &X1\_X2\_X4& 1.4576 &	0.7059	&	                         &             & \\
X3 &	2.4550 &	0.0044       & X1\_X4	& 2.1645 & 0.0653     &X1\_X3\_X4&	1.4473 & 0.7171	&		                     &             & \\
X4 &	2.4529 &0.0065         & X2\_X3	& 2,3800 &	0.0739    &X2\_X3\_X4&	1.2313 & 1.1455	&	                          &            &  \\
	 &                 &              	  & X2\_X4	& 2.3800	& 0.0728	  &				   &              &              &                           &            &  \\
	 &		             &                  & X3\_X4	& 2.3768 & 0.0760	  &				   &              &              &                           &            &  \\\hline

\end{tabular}%
}
\caption{Experiment with $Y=X_1+\sin(2\pi (X_2+X_3+X_4))+N(0,1)/10$ and $N=10,000$. Each categorized 1-features has 12 bins, so a $k$-feature has $(12)^k$ $k$D hypercubes.}
\label{summary2}
\end{table}

\begin{table}[]
\resizebox{\textwidth}{!}{%
\begin{tabular}{llllllllllll}\hline
1Feature & CE	& CE-drop	 &2Feature   &CE	      & SCE-drop &3Feature	   & CE	       & SCE-drop	& 4Feature	&CE  & SCE-drop  \\ \hline
X1 &	2.1873 &	0.2572       & X1\_X2	& 1.5863 &	0.6010    &X1\_X2\_X3&	0.3657 &	1.2022	& X1\_X2\_X3\_X4 &	0.0207	& 0.2947\\
X2 &	2.3945 & 0.0500        & X1\_X3	& 1.5679 & 0.6193    &X1\_X2\_X4&	0.3155 &	1.2601	&	                         &             & \\
X3 &	2.3789 &	0.0655       & X1\_X4	& 1.5757 & 0.6116     &X1\_X3\_X4&	0.3258 &  1.2421	&		                     &             & \\
X4 &	2.3819 & 0.0625        & X2\_X3	& 1.6502 &	0.7286    &X2\_X3\_X4&	0.3553 &  1.2718	&	                          &            &  \\
	 &                 &              	  & X2\_X4	& 1.6272	& 0.7547	  &				   &              &              &                           &            &  \\
	 &		             &                  & X3\_X4	& 1.6387 & 0.7402	  &				   &              &              &                           &            &  \\\hline

\end{tabular}%
}
\caption{Experiment with $Y=X_1+\sin(2\pi (X_2+X_3+X_4))+N(0,1)/10$ and $N=1000$. Each categorized 1-features has 12 bins, so a $k$-feature has $(12)^k$ $k$D hypercubes.}
\label{summary3}
\end{table}	

The answer to the above question is negative. That is, somehow we can escape from the curse of dimensionality in our pursuit of high order major factor. Here we demonstrate a way of escaping. We perform K-means clustering on the 3D data points of $(X_2, X_3, X_4)$ with 12, 36, 72 and 144 clusters, with which we build a new covariate feature $X_{234}$. The CEs of $X_{234}$ with respect to the four corresponding contingency tables are reported in Table~\ref{summary4} for $Y$ being categorized with 12 and 32 categories (clusters) via K-means. On the case of 12 clusters on $Y$, we see that CE of $X_{234}$ is increasing from 20 to 60 times of standard deviation (sd) away from the mean of CE of $X^{\varepsilon}_{234}$ as the numbers of clusters of $X_{234}$ increasing from 12 to 144. We observe similar evidence on the case of having $32$ categories on $Y$.

That is, we can confirm $X_{234}$ as a new order-1 major factor, which is a condensed version of $(X_2, X_3, X_4)$. Therefore, we should also claim that $(X_2, X_3, X_4)$ is indeed an order-3 major factor. This is an important and significant demonstration that we can be sure about the presence of high order major factors even when the sample size is relatively low, that is, the curse of dimensionality is escapable.
			
\begin{table}[]
\centering
%\resizebox{\textwidth}{!}{%
\begin{tabular}{|l|llll|}\hline
$Y$'s $\#$ &$X_{234}$'s $\#$ & $H[Y|X_{234}]$	& mean of $H[Y|X^{\varepsilon}_{234}]$ & $95\%$ CR of $H[Y|X^{\varepsilon}_{234}]$\\ \hline
\multirow{4}{*}{12}&12 & 2.345 & 2.394 & [2.393, 2.396]\\
&36 & 2.039 & 2.195 & [2.192, 2.198]\\
&72 & 1.783 & 1.981 & [1.978, 1.984]\\
&144 & 1.409 & 1.652  &[1.648, 1.655] \\\hline
\multirow{4}{*}{32}&12 & 3.141 & 3.192 &[3.190, 3.194]\\
&36 &  2.651 & 2.790 & [2.787, 2.794]\\
&72 & 2.180 & 2.385 & [2.382, 2.388]\\
&144 & 1.720 & 1.888 & [1.885, 1.892]\\\hline

\end{tabular}%
%}
\caption{Exploring the presence of $X_{234}$ as an order-3 major factor of $Y=X_1+\sin(2\pi (X_2+X_3+X_4))+N(0,1)/10$ with $N=1000$ with respect to 2 and 4 choices of numbers of clusters of $Y$ and $X_{234}$, respectively. The confidence intervals are calculated based on 100 simulations.}
\label{summary4}
\end{table}

Further, by contrasting Table~\ref{summary4} with Table~\ref{summary3}, the biases of mutual information estimates indeed can be managed by reducing the large number of bins, cells or hypercubes on the covariate side. That is, a small number of clusters can be derived via a clustering approach of choice.

\section{Examples with complex Re-Co dynamics with dependent covariate features.}
In this section, we conduct one experimental Re-Co dynamics defined by linear structures with slightly dependent covariate features as specified below. That is, this experiment is in the classic linear regression domain. However, there are two twists included in this experiment. The first twist is that there exist two almost-colinearity 3D hyper-planes pertaining to two triplets of covariate features. The second twist is that, when a continuous measurement data type is altered into a categorical one, we understand that we discard very fine scale information of measurements often together with some degrees of ordinal relational information. Nevertheless, this act of investment by sacrificing some information in data is necessary for carrying out our CE computations in its quest for critical authentic information content contained in data. On the other hand, it is natural to as the following question: When linear regression analysis is applied to such a categorized data set, do we naturally expect its conclusions from such an analysis could be close to the true linear structure?

In this section, we investigate the aforementioned two twists in order to understand the general effects of dependence on conditional entropy evaluations, and we also address the above question. The particular focuses are placed on issues linking to validity of Information Theoretical measurements and their reliability evaluations. We would like to demonstrate the comparisons between classical statistics and CEDA's major factor selection upon the quests into Re-Co dynamics.

\subsection{[Example-6]: From dependency induced complications to reality.}
Consider a Re-Co dynamics defined by linear structures with slightly dependent covariate-features:
\begin{eqnarray*}
Y&=&X_1+X_2+X_3+N(0,1)/10,\\
X_6&=& (X_1+X_2+X_3+X_4+X_5+N(0,1)/10)/3,\\
&&(X_1,.., X_5, X_7,...,X_{10})\sim N(\tilde{0}, \Sigma),\\
\Sigma[i,i] &=&1, \Sigma[i,j]=0.2, \; i\neq j, \; i, j \in\{1,.., 9\}.
\end{eqnarray*}
where $\Sigma$ is a $9\times 9$ covariance matrix (not including $X_6$). Features $\{X_7, X_8, X_9, X_{10}\}$ play the roles of unrelated, but dependent noise. The design of this Example-6 is to have a seemingly dominant order-1 major factor candidate: feature $X_6$. We want to explore whether we could discover the true structure underlying the RE-Co dynamics that is a collection of 3 order-1 major factors: $\{X_1, X_2, X_3\}$, or not. Also we would like to see what realistic computational issues are generated from the dependency among all covariate features.

One million of 11dim data points are simulated and collected as the data set. We apply our CE computations by having all 1D covariate features and the response feature are categorized to have 22 bins via the same scheme used in previous section. CEs are calculated for all possible feature-sets via the contingency table platform. For expositional purpose, we only report 10 CE-values for 10 key characteristic feature-sets across 1-feature to 6-feature settings in Table~\ref{tableL02}. The summary of our findings based major factor selections are reported below.

\begin{description}
\item[1.] On 1-feature setting, $X_6$ has the lowest CE and members of $\{X_1, X_2, X_3\}$ are in the second tier by having the median tier of CEs, while the rest of covariate features are in the 3rd tier having the highest CEs. Therefore, each member of $\{X_1, X_2, X_3, X_6\}$ is a potential order-1 major factor candidate. It is noted that, though $H[Y]=3.0316$ in the 0-feature setting, it is more proper to use $H^{(1)}[Y]=H[Y|X_{10}]=2.9883$ on 1-feature setting due to the contingency tables' dimension-change from $1\times 22$ to $22\times 22$, as we have argued in the previous two sections.
\item[2.] On 2-feature setting, we take $H^{(2)}[Y]=H[Y|X_4,X_7]=2.9523$ and calculate the CE-drop of $(X_4,X_6)=2.9523-2.1321=0.8202$ and CE-drop of $X_6$ as $H^{(2)}[Y]-H[Y|X_6,X_7]=2.9523-2.3309=0.6214$. Since the CE-drop of $X_4$ is basically zero. So we know that $X_6$ and $X_4$ are potentially conditional dependent given $Y$, so are $X_6$ and $X_5$.  Likewise, we calculated CE-drops of $(X_6,X_1)$ and $X_1$ as $0.7084$ and $0.2513$. Thus, the CE-drop of $(X_6,X_1)$ is smaller than the sum of CE-drops of $X_6$ and $X_1$. This is the first evidence that $X_6$ and any individual members of $\{X_1, X_2, X_3\}$ can not be order-1 major factors, simultaneously.

In contrast, the CE-drop of $(X_1,X_2)$ is calculated as $0.6338$, which is only slightly larger than the sum of CE-drops of $X_1$ and $X_2$: $0.5026$. This evidence of so-called ecological effect indicates that $X_1$ and $X_2$ are not significantly conditional dependent, but they can be order-1 major factors simultaneously. Likewise for $X_1$ and $X_3$ and $X_2$ and $X_3$.
\item[3.] On 3-feature setting, we take $H^{(3)}[Y]=H[Y|X_7,X_8, X_9]=2.8139$ and calculate the CE-drops of $(X_1, X_2, X_3)$ and $(X_4, X_5, X_6)$ as: $2.0596$ and $1.7393$, respectively. Though these two CE-drops are more than 3 times of the sums of individual CE-drops of these two triplets, which are $0.6870$ and $0.5567$, respectively, we do not claim that the two triplets $(X_1, X_2, X_3)$ and $(X_4, X_5, X_6)$ are potential candidates of order-3 major factors. Since there is no conditional dependency claims among members of these triplets in the 2-feature setting. However, we claim that $(X_1, X_2, X_3)$ is the chief collection of 3 order-1 major factors, while $(X_4, X_5, X_6)$ is an alternative collection of 3 order-1 major factors.
\item[4.] On 4-feature setting, we take $H^{(4)}[Y]=H[Y|X_7,X_8, X_9, X_{10}]=1.6278$, which is significantly smaller than $H^{(3)}[Y]$. As expected, this is an evidence of effect of curse of dimensionality. Since the averaged cell count is less than 1 in this setting. Therefore, we can not make any structural claims here. (It is also reasonable to expect that, if the number of bins is reduced to 10, the 4-feature setting might yield stable evaluations of mutual information.)
\item[5.] On 5-feature and 6-feature settings, no creditable claims can be made due to curse of dimensionality.
\end{description}

\begin{figure}
 \centering
  \includegraphics[width=5in]{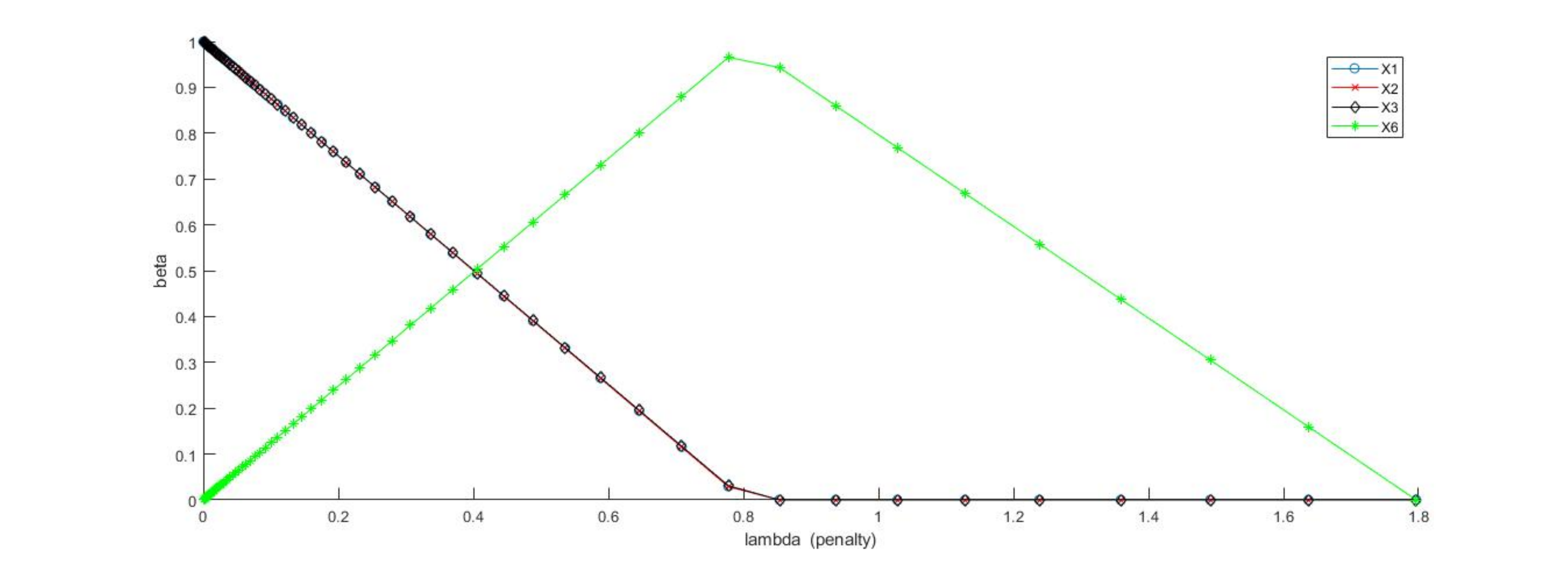}
 \caption{Results of parameters in Example-6 via LASSO with respect to a spectrum of $\lambda$ penalty values. The three cures of $X_1$, $X_2$ and $X_3$ are completely overlapping with each other.}
 \label{fig:lassodep}
 \end{figure}

Our conclusion in the 3-feature setting: the chief collection of order-1 major factors $\{(X_1, X_2, X_3)\}$ and one secondarily alternative collection $\{(X_4, X_5, X_6)\}$, is a unusual, but precise statement. This statement is in sharp contrast with classic regression analysis. For instance, for comparison purpose, we perform LASSO regressions, which is specified in the following Lagrangian form:
\[
\min_{\beta \in R^{11}}\{\| Y-X\beta\|^2_2+\lambda \|\beta\|_1\}.
\]
As shown in Figure~\ref{fig:lassodep}, the joint presence of $\{X_1, X_2, X_3, X_6\}$ are seen for all $\lambda$ falling within $(0, 0.8)$. Specifically, the observed pattern is that parameters of members of $\{X_1, X_2, X_3\}$ are linearly decreasing from 1, while parameter of $X_6$ is increasing from 0 also linearly. Such linearity is primarily due to the penalty $\lambda$. All such trajectories of $beta$ are not correct for the Re-Co dynamics except when $\lambda=0$, which only reports the result regarding $\{X_1, X_2, X_3\}$, but not $(X_4, X_5, X_6)$.

\begin{table}[]
\resizebox{\textwidth}{!}{%
\begin{tabular}{llllllllllll}\hline
1Feature & CE	 &2Feature   &CE	  &3Feature	   & CE	   	& 4Feature	&CE   & 5-feature      & CE     & 6-feature & CE\\ \hline
X6&2.3351&X4\_X6&2.1321&X1\_X2\_X3&0.7543&X1\_X2\_X3\_X7&0.5602&X1\_X2\_X3\_X7\_X8&0.1020&X1\_X2\_X3\_X7\_X8\_X9 &0.0065\\
X3&2.7295&X1\_X6&2.2439&X4\_X5\_X6&1.0746&X1\_X2\_X3\_X6&0.6201&X1\_X2\_X3\_X6\_X9&0.1723&X1\_X2\_X3\_X6\_X7\_X8 &0.0132 \\
X1&2.7308&X1\_X2&2.3184&X1\_X2\_X6&2.0049&X4\_X5\_X6\_X8&0.8789&X1\_X7\_X8\_X9\_X10&0.2255&X1\_X4\_X5\_X7\_X8\_X9 &0.0150\\
X2&2.7310&X6\_X7&2.3309&X1\_X4\_X6&2.0239&X1\_X4\_X5\_X6&0.8965&X4\_X5\_X6\_X8\_X9&0.2355& X1\_X2\_X3\_X5\_X6\_X8 &0.0202\\
X9&2.9879&X3\_X7&2.7010&X4\_X6\_X7&2.0771&X2\_X3\_X5\_X7&1.4054&X1\_X4\_X5\_X6\_X8&0.2681& X2\_X3\_X6\_X7\_X8\_X9 &0.0211\\
X8&2.9880&X3\_X4&2.7012&X3\_X6\_X9&2.1765&X4\_X6\_X7\_X9&1.4468&X5\_X6\_X7\_X8\_X9&0.2719& X4\_X5\_X6\_X7\_X8\_X9 &0.0240\\
X7&2.9882& X7\_X8&2.9516&X1\_X2\_X7&2.2328&X6\_X7\_X8\_X9&1.4605&X2\_X5\_X6\_X8\_X9&0.3022& X1\_X4\_X5\_X6\_X7\_X8 &0.0280 \\
X4&2.9882&X5\_X7&2.9520&X6\_X7\_X8&2.2572&X1\_X6\_X8\_X9&1.4752&X1\_X4\_X6\_X7\_X8&0.3035& X1\_X2\_X5\_X6\_X8\_X9 &0.0280\\
X5&2.9883&X4\_X5&2.9522&X1\_X7\_X9&2.5849&X1\_X7\_X8\_X9&1.5458&X1\_X2\_X4\_X5\_X6&0.3236& X1\_X2\_X4\_X5\_X6\_X8 &0.0329\\
X10&2.9883&X4\_X7&2.9523&X7\_X8\_X9&2.8139&X7\_X8\_X9\_X10&1.6278&X1\_X2\_X5\_X6\_X9&0.3427& X1\_X2\_X3\_X4\_X5\_X6 &0.0584\\
\hline
\end{tabular}%
}
\caption{Example-6 with  $N=10^6$. Each categorized 1-features has 22 bins, so a $k$-feature has $(22)^k$ $k$D hypercubes.}
\label{tableL02}
\end{table}	

We conclude that, though the LASSO with manmade penalty constraints seemingly coupled with some desirable interpretations, its optimization protocol clearly can not handle a landscape having two equally probable "deep-wells". In sharp contrast, our major factor selection protocol has no problems at all on identifying and confirming two collections of three order-1 major factors, and these two collections can not co-exist. This result is reiterated in the next subsection as well. This capability is the chief merit of employing Information Theoretical measures in major factor selection.

Further, we conduct the least squared estimation based on all categorized data, and report the results in Table~\ref{catindep}. We can see that the results of estimations give rise to mixed-up and wrong linear structures. That is, the categorizing scheme, which heterogeneously alters locations and scales of original data, has indeed destroyed data's intrinsic characteristics. From this perspective, we understand that the categorical nature of data is suit for Information Theoretical Measures, but not for linear regression models and its variants.

\begin{table}[]
\centering
\begin{tabular}{lllll}\hline
&Estimate&Std. Error & t value & $Pr(>|t|)$\\\hline
(intercept)&-0.776&0.013&-59.57&0.000\\
X1&0.334&0.004&819.68&0.000\\
X2&0.334&0.004&820.21&0.000\\
X3&0.334&0.004&820.05&0.000\\
X4&-0.232&0.004&-568.08&0.000\\
X5&-0.231&0.004&-566.27&0.000\\
X6&0.528&0.008&624.12&0.000\\
X7&-0.0002&0.001&-0.94&0.3462\\
X8&0.0001	&0.001&0.56&0.5735\\
X9&-0.0002&0.001&-1.40&0.1622\\
X10&-0.0002&0.001&-1.13&0.2567\\
\hline
\end{tabular}
\caption{Results of parameters in linear regression with categorized data.}
\label{catindep}
\end{table}

\subsection{Escaping from the curse of dimensionality}
In Example-6, the 6-feature setting, the feature-set $\{(X_1, X_2, X_3, X_4, X_5, X_6)\}$ achieves the largest CE among all possible feature-sets, which is at least 7 times of CE of $\{(X_1, X_2, X_3, X_7, X_8, X_9)\}$. Such comparisons are invalid due to finite sample phenomenon or curse of dimensionality. Since there are more than $1.408$ billions ($(22)^{7}$) 7D hypercubes for just one million data points. How can we escape from the potential effects of curse of dimensionality on estimations of CEs of $\{(X_1, X_2, X_3, X_4, X_5, X_6)\}$  and $\{(X_1, X_2, X_3, X_7, X_8, X_9)\}$?

Again, we apply the simple approach via K-means clustering algorithm. We first apply K-means algorithm to have 22 clusters based on one million of 3D data points of $\{(X_1, X_2, X_3)\}$, $\{(X_4, X_5, X_6)\}$ and $\{(X_7, X_8, X_9)\}$ , respectively. We specifically denote these three categorical variables as $X_{123}$, $X_{456}$ and $X_{789}$, respectively. Upon these three new covariate variables, we calculate CEs (of $Y$) under 1-feature and 2-feature settings, see Table~\ref{expL020}.  We consistently confirm that $X_{123}$ and $X_{456}$ are not conditionally dependent given $Y$. Therefore, the two feature triplets $(X_1, X_2, X_3)$ and $(X_4, X_5, X_6)$ are two separate chief and alternative collections of three order-1 major factors.

 \begin{table}[]
\centering
\begin{tabular}{|l|llll|}\hline
                experiments      & 1-feature & CE & 2-feature & CE \\\hline
\multirow{3}{*}{L0.2}  & $X_{123}$    & 1.9317  &$X_{123}$\_$X_{456}$& 1.8206 \\
                                     & $X_{456}$   &  2.4734   & $X_{123}$\_$X_{789}$  & 1.9195 \\
                                     & $X_{789}$    & 2.9450   & $X_{456}$\_$X_{789}$    & 2.4555 \\\hline
\end{tabular}%
%}
\caption{Escaping from the curse of dimensionality in Example-6.}
\label{expL020}
\end{table}

\section{Conclusion}
The most fundamental concept underlying all practical guidelines we have learned from the series of increasingly complex examples in this paper is: the comparability of evaluations of conditional entropy and mutual information critically rests on the equality of dimensions of contingency tables, where these evaluations are carried out. Based on this comparability concept, the focal goal of data analysis is then rephrased in terms of [C1: confirmable] criterion regrading presence and absence of major factors underlying a designated Re-Co dynamics. In other words, it is absolutely essential to note that there is no need of precise theoretical information measurements in real data analysis. Such [C1: confirmable] criterion pertaining to discovery of major factor subsequently promotes all practical guidelines being centered around the task of confirming and debunking an existential collection of major factors of various orders. Since presence and absence of such an existential collection of major factors indeed manifest the data's authentic information content. Hence, from data's information content perspective, the task of data analysis as one whole is translated into the single issue of major factor selection.

Further, all practical guidelines on evaluating mutual information, in particular, for our major factor selection protocol are by and large recognized for ascertaining the [C1: confirmable] criterion against the effects of curse of dimensionality or finite sample phenomenon. Practically, we learn to be sensitively aware of dangers of having low cell-counts in potentially occupied cells when evaluating entropy measures. We also develop clustering-based approaches to lessen the effect of curse of dimensionality. After learning all these practical guidelines, we are confident in our applications of our major factor selection protocol and related Categorical Exploratory Data Analysis (CEDA) techniques on analyzing real-world structured data sets.

In many scientific fields, like biology, medicine, psychology and social sciences, many measurements are not always precisely metric. Even within a metric system, a continuous measurement is often grouped and converted into a discrete or even ordinal data format. That is, very fine scale details of a data point is likely given up because it is either too costly to measure, or even can't be measured, or needs to be discarded for practical computational  considerations. Therefore, any structured data set is likely consisting of some features having incomparable measurement scales and some features have no scales at all. How to analyze such a data set in a coherent fashion is not at all a simple task. CEDA is a data analysis designed to be coherent with all features' measurements. So, CEDA and its major factor selection protocol are developed to indeed embrace the ideal concept: Each single feature must allowed to contribute its own authentic information locally, and then to congregate and weave patterns that reveal heterogeneity on global, median and fine scales levels.

To facilitate and carry out such a fundamental concept of data analysis, CEDA is exclusively resting on one simple fact: All data-types are embedded with the categorical nature. So all pieces of local information derived from all categorical or categorized features must be comparable. All these information pieces can be then woven together for the multiscale heterogeneity. By doing so, there are no man-made assumptions or structures needed in CEDA. So, information brought out by CEDA is authentic. That is, we can be free from the danger of generating misinformation via data analysis involving unrealistic assumptions or structures.

To achieve aforementioned goals of CEDA via carrying out our major factor selection protocol, we definitely need stable and creditable evaluations of conditional entropy and mutual information underlying any targeted Re-Co dynamics of interest. That is why these practical guidelines learned in this paper become essential and significant. On the other hand, these practical guidelines also reveal aspects of flexibility and capability of CEDA and its major factor selection in helping scientists to extract intelligence from their own data sets.

As the final remark, we clearly demonstrate in this paper that, by reframing many key statistical topics in one Re-Co dynamics framework, CEDA and its major factor selection protocol not only can resolve the original data analysis tasks, but also more importantly can shed authentic lights on issues related to widely expanded frameworks containing the original statistical topics. This capability manifests the capability of CEDA and its major factor selection protocol for truly accommodating and resolving real-world scientific problems.

Finally, we conclude that learned practical guidelines for evaluating CE and $I[Re; Co] $ would allow scientists to effectively carry out CEDA and its major factor selection protocol to extract data's visible and authentic information content, which is taken as the ultimate goal of data analysis.

%%%%%%%%%%%%%%%%%%%%%%%%%%%%%%%%%%%%%%%%%%

%%%%%%%%%%%%%%%%%%%%%%%%%%%%%%%%%%%%%%%%%%
\vspace{6pt}

\conflictsofinterest{The authors declare no conflict of interest.} 

%%%%%%%%%%%%%%%%%%%%%%%%%%%%%%%%%%%%%%%%%%

%%%%%%%%%%%%%%%%%%%%%%%%%%%%%%%%%%%%%%%%%%
%% Optional

%%%%%%%%%%%%%%%%%%%%%%%%%%%%%%%%%%%%%%%%%%
\begin{adjustwidth}{-\extralength}{0cm}

\reftitle{References}

\end{adjustwidth}
\end{document}